\documentclass[onecolumn]{elsart3p}
\usepackage{ifpdf}

\usepackage{graphicx}
\usepackage{amssymb}
\usepackage{amsmath}
\usepackage{cite}
\usepackage{color}

\journal{Physica D: Nonlinear Phenomena}

\newcommand{\sech}{\;{\rm sech}}
\newcommand{\csech}{\;{\rm csech}}

\newcommand{\ExpIntegralEi}{\;{\rm ExpIntegralEi}}
\newcommand{\Erfi}{\;{\rm Erfi}}
\begin{document}

\begin{frontmatter}

\title{Bright and dark solitons in a quasi 1D Bose-Einstein condensates modelled by 1D Gross-Pitaevskii equation with time-dependent parameters}
\author[cnld]{S. Rajendran},
\author[phys]{P. Muruganandam}, and
\author[cnld]{M. Lakshmanan}

\address[cnld]{Centre for Nonlinear Dynamics, Bharathidasan University, Tiruchirappalli -- 620024, Tamil Nadu, India}
\address[phys]{School of Physics, Bharathidasan University, Tiruchirappalli -- 620024, Tamil Nadu, India}

\begin{abstract}
We investigate the exact bright and dark solitary wave solutions of an effective one dimensional (1D) Bose-Einstein condensate (BEC) by assuming that the interaction energy is much less than the kinetic energy in the transverse direction. In particular, following the earlier works in the literature P\'erez-Garc\'ia et~al.~\cite{Perez2006}, Serkin et~al.~\cite{Serkin2007},  G\"urses ~\cite{Gurses2007} and Kundu ~\cite{Kundu2009}, we point out that the effective 1D equation resulting from the Gross-Pitaevskii (GP) equation can be transformed into the standard soliton (bright/dark) possessing, completely integrable 1D nonlinear Schr\"odinger (NLS) equation by effecting a change of variables of the coordinates and the wave function.  We consider both confining and expulsive harmonic trap potentials separately and treat the atomic scattering length, gain/loss term and trap frequency as the experimental control parameters by modulating them as a function of time. In the case when the trap frequency is kept constant, we show the existence of different kinds of soliton solutions, such as the periodic oscillating solitons, collapse and revival of condensate, snake-like solitons, stable solitons, soliton growth and decay and formation of two-soliton like bound state, as the atomic scattering length and gain/loss term are varied. However when the trap frequency is also modulated, we show the phenomena of collapse and revival of two-soliton like bound state formation of the condensate for double modulated periodic potential and bright and dark solitons for step-wise modulated potentials.

\end{abstract}

\begin{keyword}

Bose-Einstein condensates; Gross-Pitaevskii equation; Nonlinear
Schr\"odinger equation; Soliton solutions.

\PACS 05.45.-a, 03.75.Lm, 03.75.Nt, 67.85.Jk
\end{keyword}

\end{frontmatter}

\maketitle

\section{Introduction}

Experimental realization of trapped Bose-Einstein condensates (BEC) in alkali metal atoms has triggered immense interest in understanding the various properties of ultra cold matter \cite{Davis1995,Ensher1996}. These include four wave mixing, formation of vortices, bright solitons, dark solitons, gap solitons, interference patterns and domain walls in binary BECs~\cite{Rosenbusch2002, Strecker2002, Khaykovich2002, Cornish2006, Burger1999, Anderson2001,Denschlag2000,Bongs2003,Ginsberg2005,Eiermann2004,Liu2000,Malomed2004}.
An interesting dynamical feature in the context of Bose-Einstein condensation is the formation of matter wave solitons~\cite{Ruprecht1995,Perez1998,Burger1999,Anderson2001,Denschlag2000,%
Busch2000,Trombettoni2001,Khaykovich2002,Strecker2002,Cornish2006, Salasnich2003,Bongs2003,Eiermann2004,Ginsberg2005}. For example, dark solitons \cite{Burger1999,Anderson2001,Denschlag2000,Bongs2003,Ginsberg2005} and gap solitons \cite{Eiermann2004} were experimentally observed in BECs with repulsive interactions, whereas bright solitons  were demonstrated in systems with attractive interactions~\cite{Khaykovich2002,Strecker2002}. The recent experiments at Heidelberg and Hamburg universities have shown the formation of dark solitons, their oscillations and interaction in single component BECs of $^{87}$ Rb atoms with confining harmonic potential~\cite{Becker2008,Weller2009,Stellmer2008}.
These experimental results have been augmented by extensive theoretical work including predictions on bright and dark matter wave solitons~\cite{Ruprecht1995,Perez1998,Busch2000}, lattice solitons \cite{Trombettoni2001}, and soliton trains \cite{Salasnich2003}.

The dynamics of a BEC at absolute zero temperature is usually described by the mean-field Gross-Pitaevskii (GP) equation~\cite{Pethick2002}, which is a generalized form of the ubiquitous NLS equation\cite{Ablowitz1991,Agrawal1995,Lakshmanan2003}, for the wave function of the condensate. The NLS equation and its generalizations are widely used in the nonlinear optics literature~\cite{Ablowitz1991,Agrawal1995,Lakshmanan2003} to describe the evolution of coherent light in a nonlinear Kerr medium, envelope dynamics of quasi-monochromatic plane wave propagation in a weakly dispersive medium, high intensity pulse propagation in optical fibers and so on.

As is well known the NLS equation in one space and one time dimension is a completely integrable soliton system~\cite{Ablowitz1991,Agrawal1995,Lakshmanan2003}: It possesses bright solitons in the focusing case, while dark solitons exist in the defocusing case. On the other hand, the GP equation which is a three space and one time dimensional nonlinear evolution equation, is not integrable in general, but can admit exact solutions only in very special cases. In general, the GP equation has to be analyzed numerically or by approximation methods.  For the cigar-shaped trap potential when the transverse linear oscillator length is much smaller than  the longitudinal length, the 3D GP equation can be reduced to the effective 1D GP equation by assuming the kinetic energy of the longitudinal excitations and the two-body interaction energy of the atoms are of the same order, while both are much less than the kinetic energy of the transverse excitations~\cite{Salasnich2002,Brazhnyi2003}. To obtain bright and dark soliton solutions, its quasi-1D version has to be analyzed essentially numerically. 

Interestingly, some of the soliton solutions in the presence of regular harmonic confinement, with appropriately tailored gain or loss, have been shown to exhibit collapse and revival phenomena, with an increase in amplitude. For example, Hulet and his co-workers observed the growth of a condensate of trapped $^{7}$Li atoms with attractive interaction and subsequent collapse \cite{Gerton2000}, and Miesner et~al. measured the rate of growth of a $^{23}$Na condensate fed by a thermal cloud \cite{miesner1998}. Such a growth of a BEC from thermal vapor was also
experimentally realized for atoms of $^{87}$Rb~\cite{Kohl2002}. Several methods have been introduced theoretically to account for the growth of BEC \cite{Gardiner1998,Lee2000,Pethick2000,Drummond2000}. The GP equation with the complex linear gain or loss term, often referred to as GP gain equation, was used to model this type of BEC dynamics by Drummond and Kheruntsyan~\cite{Drummond2000}. It has been shown that, as the condensate grows, the center of mass oscillates in the trap. The GP gain equation is of special importance in the field of atom laser, which is a device that produces an intense coherent beam of atoms by stimulated process \cite{Kneer1998,Treiber1996}. In recent years, several attempts have been made to understand the dynamics of BEC by studying the GP gain
equation~\cite{Serkin2000,Kruglov2003,Atre2006}. 

Further, there has been an increased interest during  recent times in studying the properties of BECs with time varying control parameters. In particular, investigations have been made over a decade or so to understand the properties of condensates under
\begin{enumerate}
\item[(i)] the temporal variation of atomic scattering length which can be achieved through Feshbach resonance~\cite{Moerdijk1995,Roberts1998,Stenger1999,Cornish2000},

\item[(ii)] inclusion of appropriate time-dependent gain or loss terms which can be phenomenologically incorporated to account for the interaction of atomic cloud or thermal cloud~\cite{Atre2006}, 

\item[(iii)] periodic modulation of trap frequencies\cite{Janis2005} and so on.
\end{enumerate}

Earlier studies on BECs have shown that a suitable periodic modulation of scattering length can stabilize the condensate against collapse \cite{Saito2003,Abdullaev2003,Adhikari2004}. However, there are certain occasions where a sign alternating nonlinearity and increase in the frequency of oscillations result in the acceleration of collapse~\cite{Konotop2005}. In this connection, a systematic study on the condensates under the influence of time-dependent parameters which can be achievable through experiments is of considerable importance.

 Most of the theoretical studies on the matter wave solitons in BECs have been carried out using either numerical or certain approximation (perturbation) methods~\cite{Ruprecht1995,Perez1998,Busch2000}. For example, Ruprecht et~al~\cite{Ruprecht1995} have shown the existence of bright soliton solutions in weakly interacting 1D and 2D BECs by solving the corresponding GP equations numerically. In an another context, P\'erez-Garc\'ia et~al~\cite{Perez1998} have shown the existence of bright soliton solutions to the GP equation with negative scattering length in highly asymmetric traps by using the multiple-scale approximation method, while Busch and Anglin~\cite{Busch2000} have shown the motion of dark solitons in trapped effective 1D BEC by using a multiple time scale boundary layer theory. On the other hand, there are only a few exact analytical methods available in the literature~\cite{Atre2006,Perez2006,Serkin2007,Gurses2007,Kundu2009} to deal with the GP equation. In this connection it is worth mentioning that Serkin et~al~\cite{Serkin2007} have constructed exact soliton solutions by relating the effective 1D GP equation to the compatibility condition of a Lax pair which is in turn related to that of the Lax pair of the 1D integrable NLS equation using an inverse scattering transform (IST) method while Atre et~al~\cite{Atre2006} transformed the GP equation to the unforced Duffing equation and thereby derived soliton solutions from the elliptic function solutions of the Duffing equation. Also, Perez-Garcia et~al.~\cite{Perez2006} obtain a similarity transformation connecting the NLS equation with time varying coefficients and the autonomous  cubic NLS equation. In addition, there exists several other interesting works in the literature to construct exact soliton solutions of the 1D GP equation: Specifically, for certain time-dependent atomic scattering lengths (nonlinearity) with (i)  time independent homogeneous, linear and quadratic potentials without gain/loss term~\cite{Khawaja2006}, (ii)  time independent expulsive parabolic potential without gain/loss term~\cite{Liang2005},    (iii) time independent expulsive parabolic potential  and arbitrary constant gain/loss term ~\cite{li2008}, (iv) tanh type modulated harmonic trap frequency without gain/loss term~\cite{Xue2005}, explicit growing/decaying type solutions have been constructed. In the optical soliton context, in Ref.~\cite{Wu2008} exponentially growing solutions have been constructed for either constant atomic scattering length or zero potential case with a specific tanh type dissipation term.  However, a more detailed analysis for bright and dark soliton solutions in quasi-1D BEC with time-dependent  atomic scattering length, harmonic trap frequency and gain/loss term is of considerable present interest from an experimental point of view.

In this paper we bring out a large class of bright and dark soliton solutions by simply mapping the effective 1D GP equation onto the standard NLS equation, following the earlier works in the literature~\cite{Perez2006,Serkin2007,Gurses2007,Kundu2009}. In particular we show that the effective 1D equation resulting from the Gross-Pitaevskii (GP) equation in confining as well as expulsive potentials with time varying parameters, namely the scattering length, gain/loss term and trap frequency, can be transformed into the standard soliton (bright/dark) possessing, completely integrable 1D NLS equation by effecting a change of variables of the coordinates and the wave function under certain condition, which is related to the scattering length, gain/loss term  and the trap frequency. We consider both confining and expulsive harmonic trap potentials separately and treat the atomic scattering length, gain/loss term and trap frequency as experimental control parameters by modulating them as a function of time.  We have deduced different kinds of bright/dark soliton solutions for different forms of the trap frequency by suitably tailoring the atomic scattering length and atom gain/loss term such as 
\begin{enumerate}
 \item[(i)] periodic oscillating solitons, collapse and revival of condensate, and  snake-like solitons for time independent confining harmonic potential,

 \item[(ii)] stable solitons, soliton growth and decay, and formation of two-soliton bound state for time independent expulsive harmonic potential,

\item[(iii)] Oscillatory solitons, phenomena of collapse and revival of two-soliton like bound state formation of the condensate for double modulated periodic potential and

 \item[(iv)] stable solitons, soliton growth and decay for step-wise modulated potential. 
\end{enumerate}

We also make comparisons of our results with the experiments of Khaykovich et~al~\cite{Khaykovich2002}and Strecker et~al~\cite{Strecker2002} obtained for $^7$Li atoms and for $^{87}$Rb atoms~\cite{Becker2008,Weller2009,Stellmer2008}. Further, we also compare our results with the theoretical results of Atre et~al~\cite{Atre2006} and Serkin et~al~\cite{Serkin2007}.

 It may be further noted that in experiments the matter wave solitons~\cite{Khaykovich2002,Strecker2002,Becker2008,Weller2009,Stellmer2008} have been observed within few micrometer ($\mu$m) lengths and times up to the order of few ms. In the present paper we choose the soliton parameters such that the length ($\mu$m) and time (ms) scales are well within the experimental regime. Further, we have confined ourselves to the issue of stabilizing solitons by choosing the forms of modulated atomic scattering length and atom gain/loss term so as to balance each other. We have also considered here only the two-body interactions and neglected the three-body recombination loss which will introduce a quintic term in the GP equation which does not in general reduce to exactly integrable standard NLS equation. Such a case may be studied by the numerical and approximation methods. Another interesting aspect of the temporal modulations is to look into the problem of parametric amplification, which however falls outside the scope of the present analysis  and should be dealt with separately. It may also be noted that in the context of optical systems with varying coefficients, only picosecond optical solitons have been observed~\cite{Serkin2001,serkin2002,Hernandez2005,ponomarenko2006,Zhao2008,zhao2009:epjd,zhao2009:pra}.

This paper is organized as follows. In Sec.~\ref{sec:gpe}, we discuss the method of transforming cylindrically symmetric GP equation with time varying parameters such as the atomic scattering length, gain/loss and with both confining and expulsive harmonic potentials into an effective 1D GP equation with time varying parameters. Then, we map the 1D GP equation to the standard NLS equation by making use of suitable transformations, subject to a constraint on the forms of scattering length, gain/loss term and the trap frequency. Next, in Sec.~\ref{sec:solitons} we deduce the generalized bright and dark matter wave soliton solutions of the GP equation from the soliton solutions of standard NLS equation. We show the existence of various forms of bright and dark soliton solutions, their collapse and revival, and decay and growth for suitable choices of the condensate parameters with time independent confining and expulsive harmonic potentials in Secs.~\ref{ind:con} and \ref{ind:exp}, respectively. This is followed by a demonstration of various forms of bright and dark matter wave soliton solutions for temporally modulated potentials in Sec.~\ref{de}. Finally, in Sec.~\ref{con}, we present a summary of the results and conclusions. In Appendix~\ref{app:validity}, we have shown the validity criteria of the 1D GP equation for various forms of control parameters studied in this paper. In Appendix~\ref{app:solve}, we give a brief description on the solution procedure of solving the restrictive condition for which the effective 1D GP equation can be mapped on to the standard NLS equation. We also tabulate the possible solvable choices of the modulated trap frequency forms from the Ricatti equation which specifies the restrictive condition in Appendix~\ref{app:ricatti} and provide a table of physical meaning of symbols used in the text in Appendix~\ref{app:symbol}. 

\section{Gross-Pitaevskii equation with complex gain or loss term}
\label{sec:gpe}

In the present study, we consider the GP equation with the inclusion of a complex gain or loss term of the form
\begin{align}
i \hbar  \frac{\partial \Psi({\mathbf r},t)}{\partial
t}= \left[-\frac{\hbar^2}{2m} \nabla ^2+V_{\text{trap}}({\mathbf r},t)  - \sigma U \vert \Psi \vert ^2 +i \hbar \frac{\Gamma}{2}\right] \Psi({\mathbf r},t), \label{3dGPE}
\end{align}
where $U=\frac{4 \pi \hbar^2 }{m}a_s(t)$, $a_s(t)$ is the time varying s-wave
scattering length,  $V_{\text{trap}}({\mathbf r},t) = V_0(x,y)+V_1(z,t)$, $V_0(x,y)= \frac{m}{2} \omega_{\perp}^2(x^2+y^2)$ is a cylindrical time independent harmonic trap, $V_1(z,t)= \frac{m}{2}\omega_{z}^2(t) z^2$ is a time-dependent harmonic trap along the $z$ direction which can be either confining or expulsive, $\sigma = \pm 1$ depending on whether the interaction is  attractive $(\sigma = +1)$ or repulsive $(\sigma = -1)$, and $\Gamma(t)$ is the gain/loss term which is phenomenologically incorporated to account for the interaction of atomic cloud or thermal cloud. The plus sign of the $\Gamma(t)$ term leads to the mechanism of loading external atoms (thermal clouds) into the BEC by optical pumping while the minus sign of $\Gamma(t)$ describes a BEC that is continuously depleted (loss) of atoms. If the condensate is fed by a surrounding thermal cloud, then the condensate undergoes a periodic growth and collapse. The above equation has gained much importance in recent times to understand the growth and decay of the condensate due to time-dependent parameters~\cite{Serkin2000,Kruglov2003,Atre2006}. However, one has to note that the parameters are varied sufficiently slowly so as to satisfy the adiabaticity criteria so that the wave-packet continues to evolve in the same quantum eigenstate as a function of time. A possible condition for the validity of the adiabaticity criteria~\cite{Band2002a,Band2002b} is that the characteristic time scale, say $T$, of the variation of the modulated parameters is slow enough when compared with the instantaneous nonlinear eigenvalue $E_{\text{nl}}(t)$ of the nonlinear equation at time $t$, such that $E_{\text{nl}}(t) T/(2 \pi) \gg 1$.

We also note that for the GP equation~(\ref{3dGPE}) the normalization condition in the presence of the gain/loss term $\Gamma(t)$ becomes
\begin{align}\label{norm}
\int_{-\infty}^{\infty}\vert \Psi \vert^2 d^3{\mathbf r} = N  \exp\left[\int \Gamma(t) dt\right],
\end{align}
where $N$ is number of the condensate atoms.

There are typical characteristic length scales to specify the condensate density. The confining frequencies of the harmonic trap potential ($V_{trap}$), that is $\omega_{\perp}$  and $\omega_z(t)$, set the length scale for spacial size which are the transversive oscillator length $a_{\perp}=(\frac{\hbar}{m \omega_{\perp}})^{1/2}$ and the longitudinal length  $a_z(t)=(\frac{\hbar}{m \omega_z(t)})^{1/2}$. In addition, the effective mean-field  nonlinearity introduces the length scale, namely the healing length $\zeta = 8 \pi n a_s(t)$, which  is the distance over which the kinetic energy and the interaction energy balance ~\cite{carr2000,Gorlitz2001,Brazhnyi2003}. Here $n=N /(a_{\perp}^2 a_z(t))$ is the mean density of the  condensate.

The presence of the nonlinear term in the above GP equation (\ref{3dGPE}) complicates its solution procedure. One should note that it is very difficult to obtain closed form solution of Eq.~(\ref{3dGPE}) as such. In most cases the above equation can be studied either by approximation methods or by suitable numerical methods. However, under certain circumstances, Eq.~(\ref{3dGPE}) can be reduced to a simpler form by appropriate transformation whose solutions are known. One way here is to reduce or relate the above Eq.~(\ref{3dGPE}) to the standard NLS equation in (1+1) dimensions so that the soliton solutions can be readily obtained. In the following we shall describe a method of reducing Eq.~(\ref{3dGPE}) to the NLS equation and obtaining the consequent spatiotemporal patterns.

First let us reduce the GP equation (\ref{3dGPE}) to an effective 1D equation which deals with the cigar-shaped trap potential when the transverse linear oscillator length $a_{\perp}$ satisfies the following criteria: $a_{\perp}$ is much smaller then the longitudinal length $a_z(t)$~\cite{carr2000,Gorlitz2001,Brazhnyi2003}, that is  
\begin{align}
\frac{a_{\perp}}{a_z(t)}=\sqrt{\frac{\vert \omega_z(t) \vert}{\omega_{\perp}} }\ll 1,  \,\,\, a_{\perp} \sim \zeta. 
\end{align}
The above criteria essentially means that only for an axially open trap ($a_{\perp} \ll a_z(t) $) one obtains a 1D GP equation.
In this case we assume  that the condensate wave function can be taken as
\begin{align}
\Psi({\bf r},t)= & -\frac{1}{\sqrt{2 \pi a_B }a_{\perp}}\exp\left(-i \omega_{\perp}t-\frac{x^2+y^2}{2 a_{\perp}^2} \right)
\Phi(\frac{z}{a_{\perp}},\omega_{\perp}t). \label{trans1}
\end{align}
 Here the width and length of the condensate are of the order of $a_{\perp}$ and $a_z$, respectively, and hence $n=N /(a_{\perp}^2 a_z)$ is the mean density of the  condensate. Note that the kinetic energy of the longitudinal excitations [$\sim\hbar^2/2m a_z^2(t)$] and the the two-body interaction energy of atoms [$\sim (\hbar^2 a_s/m)n$] are of the same order, while both are much less than the kinetic energy of the transverse excitations ($\sim\hbar^2/2ma_{\perp}^2$)~\cite{Salasnich2002, Brazhnyi2003}. From this, we get following criteria for the validity of the reduction of the 3D GP equation to the effective 1D GP equation in the \emph{absence} of the gain/loss term $\Gamma(t)$~\cite{Salasnich2002, Brazhnyi2003},
\begin{align}
\frac{2 a_s(t) N}{a_z(t)} \ll 1. \label{cre} 
\end{align}
However, in the presence of the gain term  in Eq.~(\ref{3dGPE}), and in view of the normalization condition~(\ref{norm}), the above criterion may be modified by replacing $N$ with $N \exp\left[\int \Gamma(t) dt\right]$ as 
\begin{align}
\frac{2 a_s(t) N \exp\left[\int \Gamma(t) dt\right]}{a_z(t)} \ll 1. \label{cre1} 
\end{align}
 Note that due to the above condition, for a given $N$, the functional form of the scattering length $a_s(t)$, gain/loss term $\Gamma(t)$ and the longitudinal(axial) length $a_z(t)$ are to be chosen appropriately as a function of time so that the above condition can be valid for sufficiently long times, which we will demonstrate in Appendix~\ref{app:validity} for various forms $a_s(t)$, $\Gamma(t)$ and $a_z(t)$ studied in this paper.%

Substituting the above form of the wave function~(\ref{trans1}) in Eq.~(\ref{3dGPE}), one obtains the effective 1D GP equation with modulational gain/loss term, atomic scattering length and trap frequency as
\begin{align}
i \Phi_t=-\frac{1}{2} \Phi_{zz}-\sigma R(t) \vert \Phi \vert ^2 \Phi
+\frac{\Omega^2(t)}{2} z^2 \Phi+i \frac{\gamma(t)}{2} \Phi, \label{1dGPE}
\end{align}
where $\Omega^2=\frac{\omega_z^2}{\omega_{\perp}^2}$,
$\gamma=\frac{\Gamma}{\omega_{\perp}}$,   $a_{\perp}=(\frac{\hbar}{m \omega_{\perp}})^{1/2}$, $R(t)=\frac{2 a_s(t)}{a_B}$ and $a_B$
is the Bohr radius. Note that in eq~(\ref{1dGPE}), the variable $z$ actually represents $z/a_{\perp}$, due to the transformation~(\ref{trans1}). Similarly $t$ stands for $\omega_{\perp} t$.  Using the relation~(\ref{trans1}), we can rewrite the above condition as
\begin{align}
\int_{-\infty}^{\infty}\vert \Phi \vert^2 d^3{\mathbf r} = \frac{ 2 N a_B}{a_{\perp}} \exp\left[\int \gamma(t) dt\right].\label{norm1}
\end{align}
In addition, one can also transform away the gain (or loss) term $\gamma(t)$ in
Eq.~(\ref{1dGPE})  by applying the following transformation~\cite{Perez1995}
\begin{align}
\Phi(z,t)= \exp\left[\int \frac{\gamma(t)}{2} dt\right] Q(z,t), \label{phi}
\end{align}
so that Eq.~(\ref{1dGPE}) can be rewritten in terms of the new variable $Q(z,t)$ as
\begin{align}
i Q_t=-\frac{1}{2} Q_{zz}- \sigma \tilde{R}(t) \vert Q \vert ^2 Q +\frac{\Omega^2(t)}{2} z^2 Q, \label{nnlse}
\end{align}
where $\tilde{R}(t)$ is a time-dependent parameter and can be written as a function of $R(t)$ and $\gamma(t)$ in the following way
\begin{align}
\tilde{R}(t)= \exp \left[ \int \gamma(t) dt \right] R(t) \label{rtilde}.
\end{align}

\subsection{Transformation to standard NLS equation}

It has been shown by G\"urses~\cite{Gurses2007,Kundu2009} recently that the above Eq.~(\ref{nnlse}) can be mapped onto
the standard NLS equation under the following,
transformation,
\begin{align}
Q = \Lambda q(Z,T), \label{qtrs}
\end{align}
where the new independent variables $T$ and $Z$ are chosen as functions of the old independent variables $t$ and $z$ as
\begin{align}
T=&G(t), \;\;\; Z=F(z,t), \label{FG}
\end{align}
while $\Lambda = \Lambda(z,t)$ is a function of $t$ and $z$.It may also noted that similar mapping has been done by Serkin et~al.~\cite{Serkin2007} and P\'erez-Garc\'{\i}a et~al.~\cite{Perez2006} using the lens transformation. Applying the above transformation (\ref{qtrs}), so that
\begin{align}
\frac{\partial}{\partial t} = G_t \frac{\partial}{\partial T}+ F_t \frac{\partial}{\partial Z},  \qquad \frac{\partial}{\partial z}= F_z \frac{\partial}{\partial Z}, \qquad \frac{\partial^2}{\partial z^2}= F_{zz} \frac{\partial}{\partial Z}+ F_z^2 \frac{\partial^2}{\partial Z^2}, \qquad (G_t= \frac{\partial G}{\partial t}, F_z= \frac{\partial F}{\partial z})
\end{align}
one can reduce
Eq.~(\ref{nnlse}) to the standard NLS equation of the form
\begin{align}
i q_{T} +q_{ZZ}+ \sigma \vert q \vert^2 q =0 \label{nlse},
\end{align}
subject to the conditions that the functions $\Lambda$, $F$, $G$, $\Omega$ and
$\tilde{R}$ should satisfy the following set of equations,
\begin{subequations}
\begin{align}
& i \Lambda_t + \frac{1}{2}\Lambda_{zz}-\frac{\Omega(t)}{2} z^2 \Lambda=0, \label{eq:cond:a}\\
& i \Lambda F_t + \frac{1}{2}\left(2 \Lambda_{z} F_z+ \Lambda F_{zz}\right)=0, \\
& G_t = \frac{F_z^2}{2} = \tilde{R} \vert \Lambda \vert ^2. \label{eq:cond:c}
\end{align}\label{eq:cond}
\end{subequations}
The above NLS equation (\ref{nlse}) is widely used in the nonlinear optics to describe the evolution of coherent light in dispersive medium, high intensity pulse propagation in optical fibers and so on, which is completely integrable equation and admits soliton, elliptic and trigonometric function solutions~\cite{Ablowitz1991,Agrawal1995,Lakshmanan2003}. 

In order to solve the unknown functions $\Lambda$, $F$, $G$ in the above equations (\ref{eq:cond}) we assume the polar form%
\begin{align}
\Lambda=r(z,t) \exp[i \theta(z,t)],\label{Lambda}
\end{align}
One can immediately check from the relations~(\ref{eq:cond:c}) that $r$ is a function of $t$ only, $r=r(t)$, since $G$ and $\tilde{R}$ are functions of $t$ only. Then from Eqs.~(\ref{eq:cond}) one can easily deduce the transformation function $\Lambda$ given by~(\ref{Lambda})  and the transformations $G(t)$ and $F(z,t)$ through the following relations (Details are given in Appendix~\ref{app:solve}):
\begin{subequations}
\begin{align}
r^2  & = 2 r_0^2  \tilde{R},\\
\theta & = -\frac{\tilde{R}_t}{2 \tilde{R}} z^2 +  2 b r_0 ^2 \tilde{R}  z- 2 b^2 r_0^4 \int \tilde{R}^2 dt, \\
F(z,t) & = 2 r_0 \tilde{R} z-4 b r_0^3 \int \tilde{R}^2 dt, \\
G(t) & = 2 r_0^2  \int \tilde{R}^2 dt,
\end{align}
\label{solv}
\end{subequations}
Here $b$, $r_0$ are arbitrary constants, and $\tilde{R}$ and $\Omega^2$ should be related by the following condition (see Appendix~\ref{app:solve}, Eq.~(\ref{sol_con:b3}) for made details)
\begin{align}
\frac{d}{dt}\left(\frac{\tilde{R}_t}{\tilde{R}}\right) -\left(\frac{\tilde{R}_t}{\tilde{R}}\right)^2-\Omega^2(t)=0, \label{con:eqn}
\end{align}%
which is a Riccati type equation with dependent variable $\frac{\tilde{R}_t}{\tilde{R}}$ and independent variable $t$. Note that here $\displaystyle \tilde{R}(t)= R(t) \exp \left[ \int \gamma(t) dt \right]= \frac{2 a_s(t)}{a_B} \exp \left[ \int \gamma(t) dt \right]$ is a quantity representing the modulated scattering length and gain/loss term and $\Omega(t) = \omega_z(t)/\omega_{\perp}$ represents the time-dependent trap frequency.  Eq.~(\ref{con:eqn}) implies a specific constraint between these quantities and for a suitable choice of them satisfying the equation~(\ref{con:eqn}), the effective 1D GP equation ~(\ref{1dGPE}) and so Eq.~(\ref{nnlse}) can be mapped onto the standard NLS equation~(\ref{nlse}). It may be noted that a transformation similar to the above transformation has also been reported by Serkin et~al. \cite{Serkin2007} by relating Eq.~(\ref{nnlse}) to the compatibility condition of  the related Lax pair and P\'erez-Garc\'{\i}a et~al. \cite{Perez2006} using the lens transformation. However, we will make use of the relations (\ref{qtrs})-(\ref{con:eqn}) as they are more straightforward.  Also we note that the transformation given in Eq. (18) is more general than the one given in Refs.~\cite{Serkin2007,Perez2006} because of the additional term proportional to the arbitrary constant $b$ in Eqs.~(18 b) and (18 c). 

\subsection{Solution to the Riccati equation~(\ref{con:eqn})}

One can consider three possible approaches to the solutions to the present Riccati equation constraint~(\ref{con:eqn}):
\begin{enumerate}
\item Given $\Omega^2(t)$, obtain possible explicit forms of $\tilde{R}(t)$ by solving (\ref{con:eqn}),
\item Given  $\tilde{R}(t)$, deduce the form of the trap frequency $\Omega(t)$ as can be written down from~(\ref{con:eqn}) straightforwardly
\item Given  $\Omega^2(t)$, obtain numerical solution for $\tilde{R}(t)$ when the Riccati equation cannot be explicitly solved for $\tilde{R}(t)$
\end{enumerate}
Firstly, there are numerous forms of $\Omega^2(t)$ for which explicit solutions for $\tilde{R}(t)$ can be obtained [see for example Ref.~\cite{Polyanin1995}] by solving the Riccati equation (\ref{con:eqn}). Most important of them are given in the form of Table~\ref{table1} in Appendix~\ref{app:ricatti}. Next, for a given form of $\tilde{R}(t)$, the corresponding form of the trap frequency $\Omega(t)$ can be straightaway calculated from~(\ref{con:eqn}) as
\begin{align}
\Omega(t)=\sqrt{\frac{d}{dt}\left(\frac{\tilde{R}_t}{\tilde{R}}\right) -\left(\frac{\tilde{R}_t}{\tilde{R}}\right)^2}. \label{con:eqn1}
\end{align}
Finally, for those forms of $\Omega^2(t)$ for which no exact solutions can be found by solving the Riccati equation~(\ref{con:eqn}), numerical solutions can always be found for $\frac{\tilde{R}_t}{\tilde{R}}$, from which $\tilde{R}$ can be calculated. One can make use of all the above three approaches to obtain the matter wave solutions.

\section{Bright and dark soliton solutions}
\label{sec:solitons}

The NLS equation (\ref{nlse}) has been known for quite some time for its novel soliton solutions. It admits Jacobian elliptic function solutions, bright and dark solitary wave solutions and trigonometric function solutions~\cite{Sulem1999,Ablowitz1991}, depending on the sign of $\sigma$. For instance, soliton solutions of the standard NLS equation~(\ref{nlse}) were given by Zakharov and Shabat by solving the Cauchy initial value problem through the IST method for $\sigma=+1$ (focusing or attractive case)~\cite{Zakharov1972} and  for $\sigma=-1$~\cite{Zakharov1973} (defocusing or repulsive case). The corresponding soliton solutions are called bright ($\sigma=+1$) and dark ($\sigma=-1$) solitons, respectively. In the following, we will make use of the one soliton solutions only and higher order soliton solutions will be discussed separately. We obtain the one soliton solutions of the effective 1D GP equation~(\ref{1dGPE}), by making use of the one soliton solutions of NLS equation~(\ref{nlse}) through the relations~(\ref{phi}), (\ref{qtrs}) and (\ref{solv}).

\subsection{Attractive (focusing) type  nonlinearity ($\sigma=+1$)}

The bright soliton solution of (\ref{nlse}) for the case of $\sigma=+1$ takes the following form~\cite{Zakharov1973} (leaving aside the unimportant phase constants)
\begin{align}
q^+(Z,T)= & a \exp\left[ i \left(\frac{c }{2 }Z + \frac{2a^2-c^2}{4}T \right) \right]
\sech \left[\frac{a}{\sqrt{2}} (Z-c T)\right]. \label{bright}
\end{align}
 Here the real parameters $a$ and $c$ correspond to the amplitude and velocity, respectively, of the bright soliton envelope and plus sign in the superscript represents the $\sigma=+1$ case. Correspondingly for the GP equation(\ref{1dGPE}), using the relations~(\ref{phi}), (\ref{qtrs}) and (\ref{bright}), the bright soliton solution can be written as

\begin{align}
\Phi^+(z,t) = &\,  \exp\left[\int \frac{\gamma(t)}{2} dt\right] Q^+(z,t) = \exp\left[\int \frac{\gamma(t)}{2} dt \right] \Lambda(z,t) q^+(Z,T). \label{Phi+}
\end{align}
Now substituting the forms of $\Lambda(z,t)$, $Z$ and $T$ from Eqs.~(\ref{FG}), (\ref{Lambda}) and (\ref{solv}) in the above Eq.~(\ref{Phi+}), we get a generalized expression for the bright soliton as
\begin{subequations}
\label{eq:1d:bright}
\begin{align}
\Phi^+(z,t) = &\, A^+(t) \sech \left[\xi^+(z,t)\right] \exp\left[i\eta^+(z,t)\right], \label{Bright} 
\end{align}
Here the time-dependent amplitude of the generalized soliton (\ref{Bright}) is given by
\begin{align}
A^+(t) = &\, a r_0 \sqrt{2 \tilde{R}(t)} \mbox{e}^{\int \frac{\gamma(t)}{2} dt} = a r_0 \sqrt{2 R(t)} \mbox{e}^{\int \gamma(t) dt}, \label{amp:bright}
\end{align}
\end{subequations}
which depends on the form of the atomic scattering length $R(t)$ and gain/loss term $\gamma(t)$, and the generalized wave variable $\xi^+$, which specifies the width and velocity of the bright soliton as a function of time and longitunumberdinal distance, is given by
\begin{subequations}
\begin{align}
\xi^+(z,t) = &\, \sqrt{2}a r_0 \tilde{R} z-\sqrt{2}a r_0^2(
c+2 b r_0)\int  \tilde{R}^2 dt = \sqrt{2}a r_0 \tilde{R}\left[ z-r_0 (
c+2 b r_0) \frac{1}{\tilde{R}}\int  \tilde{R}^2 dt \right] \label{width:bright}
\end{align}
so that the width is proportional to $1/\sqrt{\tilde{R}(t)}$ and the velocity is a function of time.
Also the generalized phase of the bright soliton is given by the expression
\begin{align}
\eta^+(z,t) = &\, \frac{-\tilde{R}_t}{2 \tilde{R}}z^2+r_0\left(c +2 b r_0\right)\tilde{R} z  - r_0^2\left[\frac{(c+2 b r_0)^2}{2}-a^2\right]\int \tilde{R}^2 dt.
\end{align}
\end{subequations}

Note that the amplitude of the generalized bright soliton $A^+(t)$ given by Eq.(\ref{amp:bright}) can vary as a function of time depending upon the forms of the scattering length $R(t)$ and gain/loss term $\gamma(t)$. Consequently one can have growth or decay or constancy (of the amplitude) of the bright soliton for appropriate choices of $R(t)$ and $\gamma(t)$. In particular for the specific choice $R(t)=\exp[-2 \int \gamma(t) dt]$, the amplitude $A^+=constant$ and the soliton amplitude remains constant in time, which has not been noted in the earlier literature~\cite{Atre2006,Serkin2007,Liang2005,Xue2005,li2008,Wu2008}. However note that the  corresponding width ($\propto  1/\sqrt{\tilde{R}}=e^{-\frac{1}{2}\int \gamma(t) dt}/ \sqrt{R(t)} $) as given (\ref{width:bright}) need not be constant which can increase or decrease as the case may be. In the following sections 4-6, we will discuss different kinds of interesting generalized soliton solutions for different forms of unmodulated/modulated harmonic potentials by suitably tailoring the atomic scattering length and gain/loss term.

Finally the normalization condition (\ref{norm1}) can be rewritten in terms of the soliton parameters as 
\begin{align}
& \int_{-\infty}^{\infty} \vert \Phi^+(z,t)\vert^2 dz = 2 \sqrt{2}a r_0 \exp\left[\int \gamma(t) dt \right] = \frac{ 2 N a_B}{a_{\perp}}\exp\left[\int \gamma(\tilde{t}) d\tilde{t}\right], \label{norm21}
\end{align}
which allows one to fix the constant $r_0$ in terms of the experimental BEC parameters as
\begin{align}
r_0=\frac{N a_B}{\sqrt{2} a a_{\perp}}.
\end{align}

\subsection{Repulsive (defocusing) type nonlinearity $(\sigma=-1)$}

On the other hand, when $\sigma=-1$, one may write the dark one soliton solution in the form~\cite{Zakharov1972}
\begin{align}
q^-(Z,T)=\frac{1}{\sqrt{2}}\left[c-2 i \beta \tanh \beta (Z-c T) \right] \,
\exp\left[ -\frac{i}{2}(c^2+4 \beta^2)  T \right], \label{dark}
\end{align}
where the parameters $\beta$, $c$ correspond to the depth and velocity of the dark soliton, respectively.  Here the minus sign in the superscript means the $\sigma=-1$ case.

Using Eqs.(\ref{phi}), (\ref{qtrs}) and (\ref{dark}),
one can write down the dark soliton solution of (\ref{1dGPE}) as
\begin{align}
\Phi^-(z,t) = &\,  \exp\left[\int \frac{\gamma(t)}{2} dt\right] Q^-(z,t) = \exp\left[\int \frac{\gamma(t)}{2} dt \right] \Lambda(z,t) q^-(Z,T) \label{Phi-}
\end{align}
Now substitute $\Lambda(z,t)$, $Z$ and $T$ from Eqs.~(\ref{FG}), (\ref{Lambda}) and (\ref{solv}) in the above Eq.~(\ref{Phi-}), and we get 
\begin{subequations}
\label{eq:1d:dark}
\begin{align}
\Phi^-(z,t) = &\,   \left\{r_0 \sqrt{\tilde{R}} \exp\left[\int \frac{\gamma(t)}{2} dt\right]\right\}  \left\{c-2 i \beta \tanh\left[ \xi^-(z,t)\right]\right\} \exp\left[i\eta^-(z,t)\right], \label{Dark} 
\end{align}
Here the generalized wave variable $\xi^-$ which specifies the width and velocity of the dark soliton as a function of time and longitudinal distance is given by the expression
\begin{align}
\xi^-(z,t) = 2 \beta r_0 \tilde{R} z-2 \beta r_0^2(c+2 b r_0)\int
\tilde{R}^2 dt = 2 \beta r_0 \tilde{R} \left[z-r_0 (c+2 b r_0)\frac{1}{\tilde{R}}\int
\tilde{R}^2 dt\right] , 
\end{align}
and the generalized wave phase of the dark soliton is expressed as
\begin{align}
\eta^-(z,t) = &\, -\frac{\tilde{R}_t}{2 \tilde{R}}z^2+2 b r_0^2\tilde{R} z  -  r_0^2\left(c^2 + 2 b^2 r_0^2+4 \beta^2\right)\int \tilde{R}^2 dt.
\end{align}
\end{subequations}

Also the depth of the generalized dark soliton is of the form
\begin{align}
A^- = &\,   \left\{2 \beta r_0 \sqrt{\tilde{R}(t)} \exp\left[\int \frac{\gamma(t)}{2} dt\right]\right\}=    \left\{ 2 \beta r_0 \sqrt{R(t)} \exp\left[\int \gamma(t) dt\right]\right\} 
\end{align}
which depends on $R(t)$ and $\gamma(t)$. As in the case of the bright soliton solution we can get stable dark soliton solutions by choosing the forms of $R(t)$ and $\gamma(t)$ to balance each other so as to achieve a constant amplitude/depth. In this case also, the width depends on $R(t)$ and $\gamma(t)$ and can increase/decrease or remain constant with time.  
Again the normalization condition (\ref{norm1}) of the dark soliton solution for the GP equation leads us to the condition
\begin{align}
\int_{-\infty}^{\infty} \left[r_0^2 \tilde{R}(c^2+ 4 \beta^2)-\vert \Phi^-(z,t)\vert^2 \right ] dz & = 4 \beta r_0 \exp \left[\int \gamma(t) dt \right] =\frac{ 2 N a_B}{a_{\perp}}\exp\left[\int \gamma(\tilde{t}) d\tilde{t}\right], \label{norm22}
\end{align}
so that the parameter $r_0$ can be fixed in this case as
\begin{align}
r_0=\frac{ N a_B}{ 2 \beta a_{\perp}}.
\end{align}

\subsection{Choice of condensate parameters}
\label{parameters}
Now making use of the forms of $\tilde{R}$ for given $\Omega^2(t)$, as presented in Table~\ref{table1} in Appendix~\ref{app:ricatti} or through a detailed numerical analysis, into the above bright or dark soliton solutions (\ref{Bright}) or (\ref{Dark}), respectively, one can analyze the nature of the BEC matter wave properties.  In Table~\ref{table1}, we list out a chosen set of $27$ different choices of $\Omega(t)$ and the corresponding forms of $\tilde R(t)$ for which the Riccati type equation (\ref{con:eqn}) exhibits closed form solutions. In the following sections, we shall analyse the nature of the bright and dark BEC solitons, their growth, decay, collapse and revival for a select set of physically interesting choices of $\Omega(t)$ and $\tilde R(t)$ from Table~\ref{table1}.

Let us first focus our attention on the case of time-independent harmonic potential for which $\Omega^2(t)$ in (\ref{1dGPE}) is a constant ($\Omega^2(t)=\Omega_0^2$). In this case, we shall identify the different soliton solutions for both confining, $\Omega_0^2 > 0$, as well as expulsive, $\Omega_0^2 < 0$, potentials for various forms of scattering length $\tilde{R}(t)$ and gain/loss term $\gamma(t)$. It is of relevance here to point out that in recent experiments on the formation of bright matter wave solitons in an ultracold $^7$Li gas,  both the confining as well as expulsive potentials have been used~\cite{Khaykovich2002,Strecker2002}. In addition, such a formation of matter wave bright soliton has been achieved by temporal modulation of atomic scattering length due to Feshbach resonance~\cite{Khaykovich2002,Strecker2002}. Further, as pointed out earlier, we note that the formation of dark solitons, their oscillations and interaction in single component BECs of $^{87}$ Rb atoms with confining harmonic potential have been demonstrated in recent experiments at Heidelberg and Hamburg universities~\cite{Becker2008,Weller2009,Stellmer2008}. Motivated by these experiments, we fix the parameters such as the trap frequencies ($\omega_{\perp}$ and $\omega_{z}$), number of atoms ($N$) and scattering length in the transverse direction, $a_{\perp}$, to confirm to the experimental values. In particular we choose the parameters as $\omega_{\perp} = 2 \pi \times 710$Hz, $N=2 \times 10^4$, and $a_{\perp}=1.4$ $\mu$m. The trap frequency in the axial direction is chosen as $\omega_z= 2 \pi \times 70$Hz for the case of the confining potential, while we choose $\omega_z= 2  \pi i \times 70$Hz for the expulsive potential. Correspondingly $\Omega_0^2$ takes the value $\Omega_0^2 = \displaystyle{\frac{\omega_z}{\omega_{\perp}}}^2 \approx 0.01$ so that $\Omega_0$ is taken as $0.1$ in our analysis in the following.  Accordingly, in the entire manuscript the unit of time is taken as milliseconds and the length is expressed in micrometers. Also it is not difficult to provide a simple estimation of the validity of the reduction of 3D GP equation to the 1D GP equation given by Eq.~(\ref{cre}) in terms of dimensional units. For example, when one considers a BEC of $^{87}$ Rb atoms with $N \approx 10^4$, for $a_z \approx 4.4\,\mu$m  Eq.~(\ref{cre}) gives the estimate $a_s(t) < 0.22$nm, which is experimentally realizable. Note that, in the presence of the gain/loss term the criteria for the validity of 1D approximation becomes (see  Eq.~(\ref{cre1})) $\frac{R(t) a_B N \sqrt{\Omega(t)}}{a_{\perp}} \exp\left[\int \gamma(t) dt\right] \ll 1$. In the present analysis, we suitably tune $R(t)$, $\Omega(t)$ and $\gamma(t)$ to satisfy the above criterion, which may be still realizable (which will be  shown in the Appendix~\ref{app:validity} for various forms).

 In the next section we shall discuss the formation of both bright and dark matter wave solitons in time independent confining harmonic potential for different choices of temporal modulation of the atomic scattering length and gain/loss term. This is followed by a demonstration of the solitons in time independent expulsive harmonic potential case in Sec.~\ref{ind:exp}. Finally, In Sec.~\ref{de}, we discuss the formation of both bright and dark matter wave solitons in time-dependent confining potential such as double modulated periodic potential and step-wise modulated potential.  

\section{Time independent confining harmonic potential}
\label{ind:con}
                            
We consider the time independent confining potential with the ratio of trap frequencies obeying the relation $\Omega^2(t)=\Omega_0^2$, where $\Omega_0$ is a constant. Using $\Omega^2(t)=\Omega_0^2$ in Eq.~(\ref{con:eqn}), we obtain a particular solution for $\tilde{R}$  as, [see Table~\ref{table1}]
\begin{align}
\tilde{R} = \sec(\Omega_0 t+ \delta). \label{eq:rtilde_1}
\end{align}
Since $\tilde{R}= R(t) \exp(\int \gamma(t) dt)$,
one can identify different types of soliton solutions of Eq. (\ref{1dGPE}) for appropriate forms of $R$ and $\gamma$. Accordingly, we will choose these forms and investigate the nature of matter waves.

\subsection{Case 1: $\gamma=0$, bright soliton with periodic oscillations in amplitude}

\begin{figure}[!ht] 
\begin{center}
\includegraphics[width=0.85\linewidth]{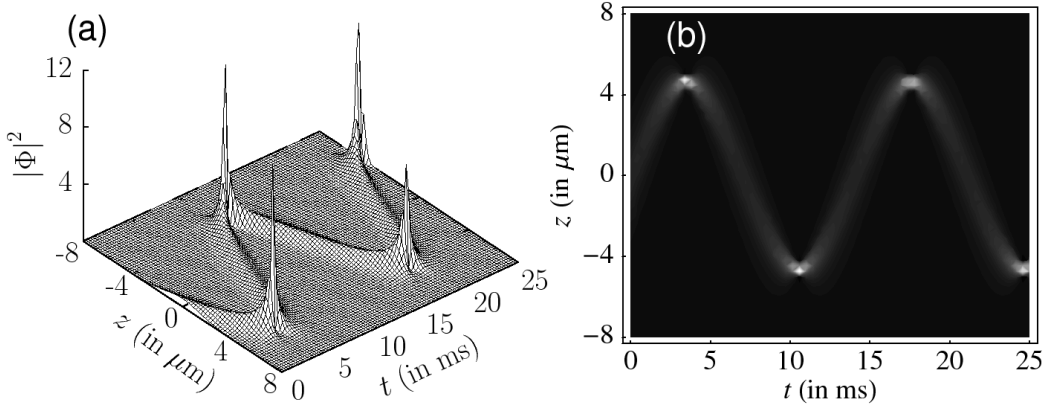}
\caption{(a) Bright soliton for the case of $\Omega^2(t)$ = $\Omega_0^2$, $\gamma=0$, and $\tilde{R}(t)=\sec (\Omega_0 t+\delta)$ and (b) the corresponding contour plot. The parameters are chosen as $a=1$, $\Omega_0=0.1$, $\delta = 0$, $b=0.1$, and $c=0.2$ in Eq.~(\ref{bright_81}). }
\label{figure1}
\end{center}
\end{figure}

When the gain term is absent in Eq.~(\ref{1dGPE}) so that for $\gamma=0$ the atomic scattering length can be written, using Eqs.~(\ref{rtilde}) and (\ref{eq:rtilde_1}), as $R=\sec(\Omega_0 t+ \delta)$. Accordingly, substituting the form (\ref{eq:rtilde_1}) into (\ref{Bright}) the magnitude of the bright soliton
solution of (\ref{1dGPE}) is given by
\begin{align}
\vert \Phi^+(z,t) \vert = &\,\left\{ \frac{N a_B}{a_{\perp}} \sqrt{\vert\sec(\Omega_0 t+\delta)\vert} \right\} \sech\left[\left(\frac{N a_B}{a_{\perp}} \sec(\Omega_0 t +\delta)\right)z-\left(\frac{N^3 a_B^3   }{a^2 a_{\perp}^3} b+\frac{N^2 a_B^2}{\sqrt{2} a a_{\perp}^2}c\right)  \frac{\tan(\Omega_0 t+\delta)}{\Omega_0}\right]. \label{bright_81}
\end{align}
It may appear that the presence of $\sec(\Omega_0 t+\delta)$ in the the amplitude of the above bright soliton will lead to singularities at finite time. However, at the same time when $\sec(\Omega_0 t+\delta) \to \infty$  the envelope term [$\sech(...)$] also goes to zero sufficiently fast so as to keep the overall amplitude finite. Note that the width of the soliton is proportional to $\sqrt{\vert\cos(\Omega_0 t+\delta)\vert}$ and hence it is varying periodically with time.
Fig.~\ref{figure1}(a) depicts the spatiotemporal pattern of condensate wave function in a confining oscillator potential for a typical choice of parameters. It is clear from Fig.~\ref{figure1}(a) that the bright matter wave
soliton wave for $R=\sec(\Omega_0 t+ \delta)$ and $\gamma=0$ shows periodic oscillation in amplitude  due to the presence of the periodic terms $\sec(\Omega_0 t+ \delta)$ and $\tan(\Omega_0 t+ \delta)$ in the right hand side of Eq.~(\ref{bright_81}). Fig.~\ref{figure1}(b) shows the periodic oscillation of the width.


In this case the corresponding dark soliton solution can be written as
\begin{align}
\vert \Phi^-(z,t) \vert = &\, \frac{N a_B}{2 \beta a_{\perp}} \sqrt{\vert\sec(\Omega_0 t+\delta)\vert} \notag \\
& \, \times \left\{c^2+4 \beta^2 \tanh^2\left[ \left(\frac{N a_B}{a_{\perp}}  \sec(\Omega_0 t +\delta)\right)z
-\left(\frac{N^3 a_B^3   }{ 2 \beta^2 a_{\perp}^3}b +\frac{N^2 a_B^2 }{2 \beta a_{\perp}^2}c\right)  \frac{\tan(\Omega_0 t+\delta)}{\Omega_0}\right]\right\}^{\frac{1}{2}}. \label{dark_81}
\end{align}
However, in the absence of a suitable gain/loss term ($\gamma=0$), the above expression becomes singular at finite values of $t$ and so is not physically interesting.

\subsection{Case 2: $\gamma=a_1 t - a_2 \sin(\kappa t)$, The collapse and revival of bright/dark solitons}
\label{case3_fig2}
A physically interesting case can be realized by choosing the gain term as $\gamma(t)=a_1 t - a_2 \sin(\kappa t)$, ($a_1$, $a_2$ and $\kappa$ are parameters), so that
\begin{align}
R(t)=  \sec(\Omega_ 0 t+\delta) \exp\left[ -\frac{a_1 t^2}{2}-\frac{ a_2 \cos(\kappa t)}{\kappa} \right] \label{r_4.1.3}.
\end{align}
The above form of $\gamma(t)$ was suggested recently by Atre, Panigrahi and
\begin{figure}[!ht] 
\begin{center}
\includegraphics[width=0.85\linewidth]{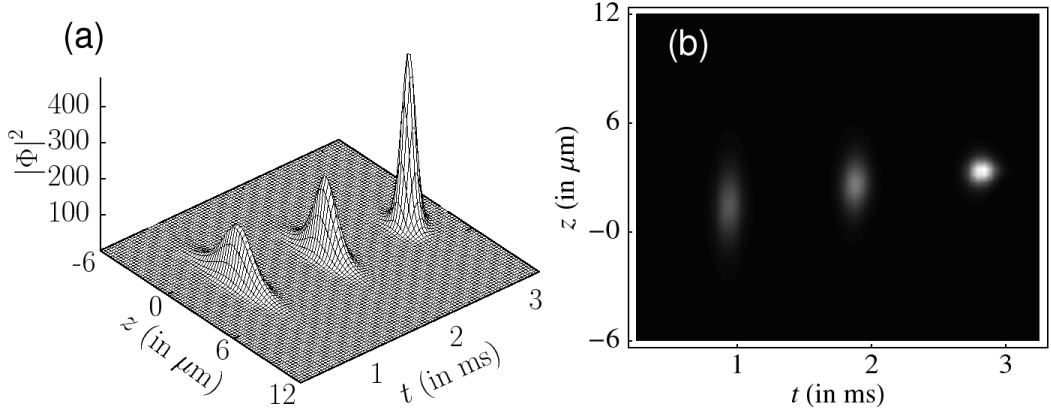}
\caption{(a) Bright matter wave soliton for $\Omega^2(t) = \Omega_0^2$, $\gamma=a_1 t - a_2 \sin(\kappa t)$, $\tilde{R}(t)=\sec (\Omega_0 t+\delta)$ with $a=\sqrt{\frac{3.125}{2}}$,  $a_1=.001$, $a_2=8.25$, $b=0$, $c=0$, $\delta=0$, $\Omega_0=0.05$, $\kappa=1.5$ in Eq.~(\ref{Bright_82}) and (b) the corresponding contour plot. Note the periodic collapse and revival of the wave with increase in amplitude over time.}
\label{figure2}
\end{center}
\end{figure} 
Agarwal~\cite{Atre2006}.
In this case, the bright soliton solution of (\ref{1dGPE}) can be written as
\begin{align}
\vert \Phi^+(z,t) \vert = &\, \frac{N a_B}{a_{\perp}} \sqrt{\vert\sec(\Omega_0 t+\delta)\vert} . \exp\left[\frac{a_1 t^2}{4}+\frac{ a_2 \cos(\kappa t)}{2\kappa} \right] \notag \\ &\,  \times\sech\left[\left(\frac{N a_B}{a_{\perp}}  \sec(\Omega_0 t +\delta)\right)z -\left(\frac{N^3 a_B^3   }{a^2 a_{\perp}^3} b+ \frac{N^2 a_B^2  }{\sqrt{2} a a_{\perp}^2}c\right)  \frac{\tan(\Omega_0 t+\delta)}{\Omega_0}\right] 
. \label{Bright_82}
\end{align}
The amplitude of the soliton is $\frac{N a_B}{a_{\perp}} \sqrt{\vert\sec(\Omega_0 t+\delta)\vert} . \exp\left[\frac{a_1 t^2}{4}+\frac{ a_2 \cos(\kappa t)}{2\kappa} \right]$  and the width is proportional to $\sqrt{\vert\cos(\Omega_0 t+\delta)\vert}$, indicating collapse and revival of the condensate. Fig.~\ref{figure2} illustrates the collapse and revival of the condensate wave function with an overall increase in the amplitude. The collapse and revival of atomic condensate and amplification [see Fig.~\ref{figure2}] through periodic exchange of atoms with the background is obviously due to the partially sinusoidal nature of the gain function, $\gamma=a_1 t - a_2 \sin(\kappa t)$.

 We also note here that, Strecker et~al. have experimentally observed the formation of the bright soliton of $^7$Li atom in quasi-1D optical trap, by magnetically tuning the interactions in a stable BEC from repulsive to attractive~\cite{Strecker2002}. Theoretically, Atre, Panigrahi and Agarwal have shown that by suitably tailoring the gain profile in the same range of the experimental parameters, the matter wave soliton exhibits dramatic collapse and revival of condensate with an increase in the amplitude~\cite{Atre2006}.

The corresponding expression for dark soliton can be written as
\begin{align}
\vert \Phi^-(z,t) \vert = &\, \frac{N a_B}{2 \beta a_{\perp}} \sqrt{\vert\sec(\Omega_0 t+\delta)\vert} \exp\left[\frac{a_1 t^2}{4}+\frac{ a_2 \cos(\kappa t)}{2\kappa}\right] \notag \\
&\, \times \left\{c^2+4 \beta^2 \tanh^2\left[  \left(\frac{N a_B}{a_{\perp}}  \sec(\Omega_0 t +\delta)\right)z
-\left(\frac{N^3 a_B^3   }{ 2 \beta^2 a_{\perp}^3}b +\frac{N^2 a_B^2 }{2 \beta a_{\perp}^2}c\right)  \frac{\tan(\Omega_0 t+\delta)}{\Omega_0}\right]\right\}^{\frac{1}{2}}. \label{dark_4.1.4}
\end{align}
The above expression, as in the case of the bright soliton~(\ref{Bright_82}), clearly shows that a periodic collapse and revival of the matter wave dark
\begin{figure}[!ht] 
\begin{center}
\includegraphics[width=0.85\linewidth]{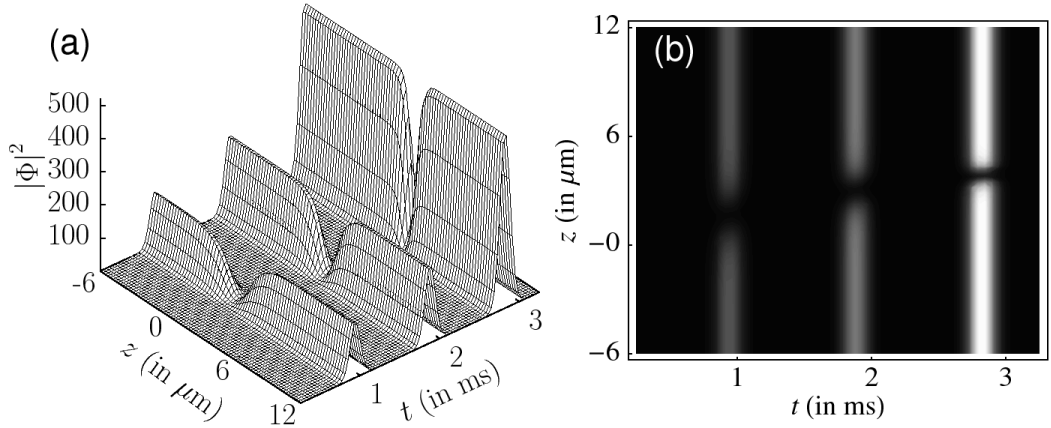}
\caption{(a) Dark matter wave soliton for $\Omega^2(t) = \Omega_0^2$, $\gamma=a_1 t - a_2 \sin(\kappa t)$, $\tilde{R}(t)=\sec (\Omega_0 t+\delta)$ with $\beta=\sqrt{\frac{3.125}{2}}$, $a_1=.001$, $a_2=8.25$, $b=0$, $c=0$, $\delta=0$, $\Omega_0=0.05$, $\kappa=1.5$ in Eq.~(\ref{dark_4.1.4}) and (b) the corresponding contour plot, again portraying a collapse and revival with increase in amplitude over time.}
\label{figure3}
\end{center}
\end{figure} 
soliton can also be realized for the choice of $R(t)$ as given by (\ref{r_4.1.3}). The dark soliton (\ref{dark_4.1.4}) is depicted  in Fig.~\ref{figure3}.

\subsection{Case 3: $\gamma=-\Omega_0 \tan( \Omega_0 t+\delta )$, snake-like bright/dark solitons}
\label{sec4:case5}

Finally we consider the case with another form of gain term as $\gamma=-\Omega_0 \tan( \Omega_0 t+\delta )$ so that $R=\sec^2 [\Omega_0 t+\delta]$ which can be obtained from Eqs.~(\ref{rtilde}) and (\ref{eq:rtilde_1}). In this case, the bright soliton solution of (\ref{1dGPE}) can be written as
\begin{align}
\vert \Phi^+(z,t) \vert = &\, \frac{N a_B}{a_{\perp}} \sech\left[\left(\frac{N a_B}{a_{\perp}}  \sec(\Omega_0 t +\delta)\right)z -\left(\frac{N^3 a_B^3   }{a^2 a_{\perp}^3} b+ \frac{N^2 a_B^2  }{\sqrt{2} a a_{\perp}^2}c\right)  \frac{\tan(\Omega_0 t+\delta)}{\Omega_0}\right]. \label{Bright_83}
\end{align}
One may note that in this case the amplitude of the soliton ($A^+ = \frac{N a_B}{a_{\perp}}$) remains constant in time. However, the corresponding width ($\sim   \sqrt{\vert\cos(\Omega_0 t+\delta)\vert}$) is varying periodically in time, leading to a snake-like effect.
\begin{figure}[!ht] 
\begin{center}
\includegraphics[width=0.85\linewidth]{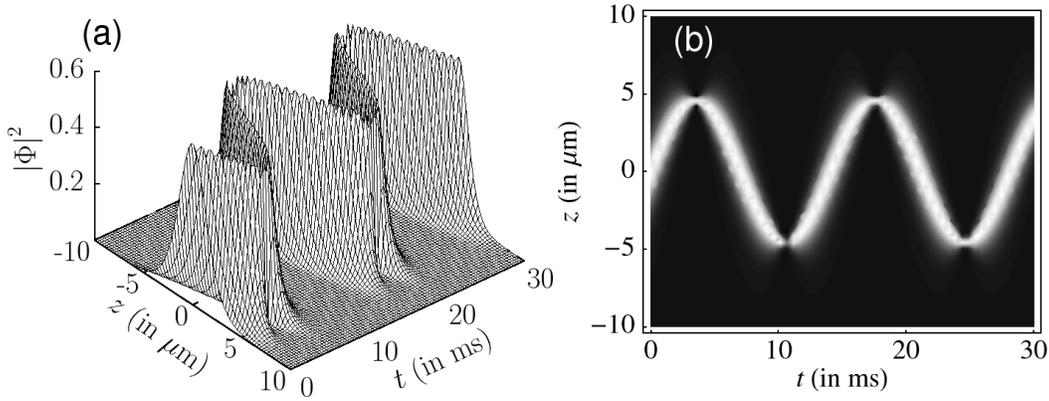}
\caption{(a) Bright soliton for $\Omega^2(t)=\Omega_0^2$ and $\gamma=-\Omega_0 \tan( \Omega_0 t+\delta )$, $\tilde{R}(t)=\sec[\Omega_0 t+\delta]$. (b) The corresponding contour plot. The other parameters are $a=1$, $b=0.1$, $c=0.5$, $\delta=0$, and $\Omega_0=0.1$ in Eq.~(\ref{Bright_83}).}
\label{fig4}
\end{center}
\end{figure} 
Figs.~\ref{fig4} shows the bright solitary wave~(\ref{Bright_83}) exhibiting a snake-like effect. It is interesting to point out that a similar snake-like effect has been demonstrated by Serkin and Hasegawa in the context of dispersion managed optical soliton with sinusoidally varying dispersion coefficient, nonlinearity, and gain or absorption term~\cite{Serkin2000}.
\begin{figure}[!ht] 
\begin{center}
\includegraphics[width=0.85\linewidth]{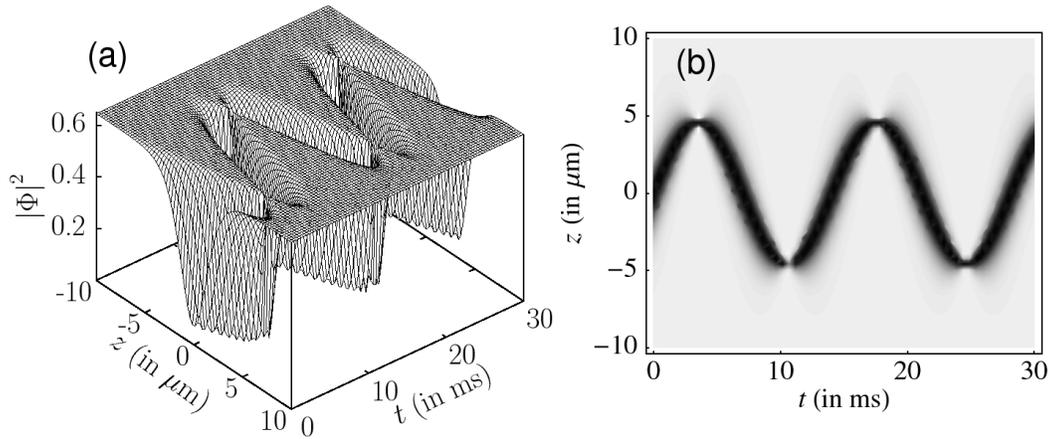}
\caption{(a) Dark soliton for $\Omega^2(t)=\Omega_0^2$ and $\gamma=-\Omega_0 \tan( \Omega_0 t+\delta )$, $\tilde{R}(t)=\sec[\Omega_0 t+\delta]$ case.  (b) The corresponding contour plot. The other parameters are $\beta=1/\sqrt{2}$,  $b=0.1$, $c=0.5$, $\delta=0$, and $\Omega_0=0.1$ in Eq.~(\ref{dark_83}).}
\label{fig5}
\end{center}
\end{figure} 


The dark soliton solution, in this case, is given by
\begin{align}
\vert \Phi^-(z,t) \vert = &  \frac{N a_B}{2 \beta a_{\perp}} \left\{c^2+4 \beta^2 \tanh^2\left[ \left(\frac{N a_B}{a_{\perp}}  \sec(\Omega_0 t +\delta)\right)z  -\left(\frac{N^3 a_B^3   }{ 2 \beta^2 a_{\perp}^3}b +\frac{N^2 a_B^2 }{2 \beta a_{\perp}^2}c\right)  \frac{\tan(\Omega_0 t+\delta)}{\Omega_0}\right]\right\}^{\frac{1}{2}}\label{dark_83}
\end{align}
As pointed out earlier in Sec.~\ref{sec4:case5} for the bright soliton, the above dark soliton (with constant amplitude and periodically varying width) also
exhibits a snake-like effect. Fig.~\ref{fig5} depicts the propagation of snake-like dark solitary wave. One may note that a similar type of snake-like  dark matter wave solitons and their collisions have been observed in the recent experiments of $^{87}$Rb condensates~\cite{Becker2008,Weller2009,Stellmer2008} with confining potential. In this context, it is quite interesting to have an exact solution of the form~(\ref{dark_83}) for the snake-like dark soliton solutions for confining potential with time varying scattering length and atom gain/loss.

\section{Expulsive potential $(\Omega^2(t)=-\Omega_0^2$ is constant$)$}
\label{ind:exp}

In this section, we consider the case of time-independent expulsive potential $\Omega^2(t) = -\Omega_0^2$. For this choice, substituting $\Omega^2(t) = -\Omega_0^2$ in the integrability condition (\ref{con:eqn}), one may identify three independent particular solutions of physical interest for $\tilde{R}$ \cite{Polyanin1995}, namely (i) $\tilde{R} = \exp(-\Omega_0 t)$, (ii) $\tilde{R} = \exp(\Omega_0 t)$, and (iii) $\tilde{R} = \sech (\Omega_0 t+ \delta)$  [see Table~\ref{table1}]. However, the stabilization of bright and dark matter wave solitons is solely dependent on the constancy of the product term $\sqrt{\tilde{R}(t)} \exp[\int \frac{\gamma(t)}{2} dt]= \sqrt{R(t)} \exp[\int \gamma(t) dt]$.
 The corresponding restriction on the parameters is called safe range parameters. When the parameters are chosen outside the safe range, the bright and dark solitons may become unstable, that is, one may witness either growth or decay,  as studied for (i) $\gamma=0$ and $R(t)=e^{\Omega_0 t}$ in Ref.~\cite{Liang2005} and (ii) $\gamma=$ constant  and $R(t)=\sech(\Omega_0 t) \,\, e^{\gamma t}$ in Ref.~\cite{li2008}. In fact, stable solitons can be formed only if we choose the form of the scattering length $R(t)$ and the gain/loss term $\gamma(t)$ as $R(t)=\exp[-2 \int \gamma(t) dt]$, so that the soliton amplitude remains constant in time. In the following, we discuss various kinds of soliton solutions of Eq. (\ref{1dGPE}) for different forms of $\gamma(t)$ and $R(t)$ both within as well as outside the safe range parameter values in all the above three cases of $\tilde{R}$.

\subsection{Case 1: $\tilde{R}=\exp(\Omega_0 t)$}
\label{+omega}
For $\tilde{R} = \exp(\Omega_0 t)$ case, we study the solution of Eq.~(\ref{1dGPE}) for both the negative (loss) and positive (gain) values of $\gamma$.  In this case, we find stable solitons for the safe range value $\gamma= -\Omega_0$ and growth of matter wave soliton for the out of safe range value $\gamma= \Omega_0$, where $\Omega_0 > 0$.

\subsubsection{$\gamma= -\Omega_0 $, stable bright/dark soliton:}
First let us consider the inclusion of depletion/loss term in (\ref{1dGPE}), that is, $\gamma = -\Omega_0 < 0$. In this case $R=\exp[2 \Omega_0 t]$, obtained from Eqs.~(\ref{rtilde}) and (\ref{eq:rtilde_1}), implying that the atomic scattering length is exponentially growing with time. Accordingly the bright soliton solution of (\ref{1dGPE}) is given by
\begin{align}
\vert \Phi^+(z,t) \vert=  &\,\frac{N a_B}{a_{\perp}}  \sech\left[\left(\frac{N a_B}{a_{\perp}}\exp(\Omega_0 t) \right)z-\left(\frac{N^3 a_B^3  }{a^2 a_{\perp}^3} b+ \frac{N^2 a_B^2  }{\sqrt{2} a a_{\perp}^2}c\right) \frac{\exp(2\Omega_0 t)}{2 \Omega_0}\right] , \label{bright3_1}
\end{align}
\begin{figure}[!ht] 
\begin{center}
\includegraphics[width=0.85\linewidth]{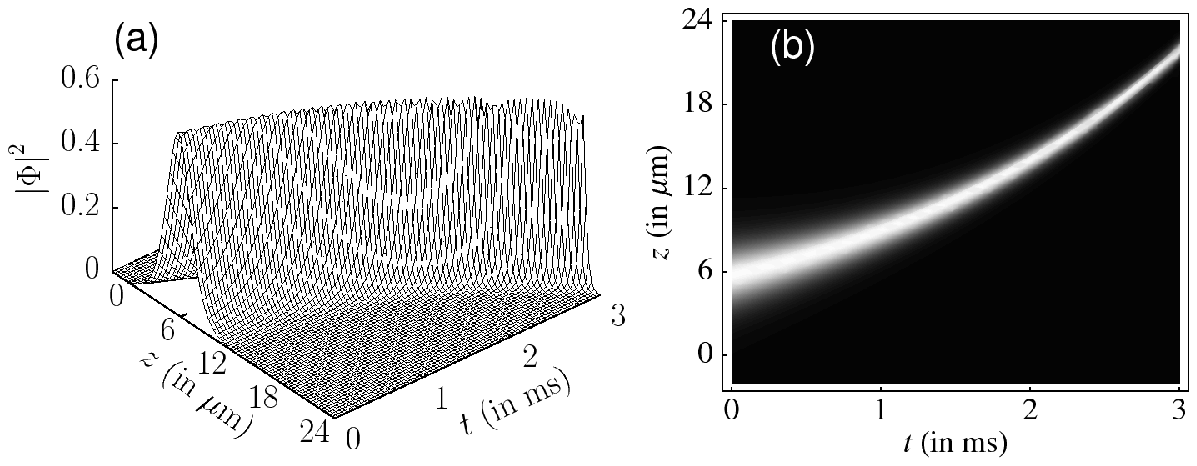}
\caption{(a)Bright soliton for $\Omega^2(t)=-\Omega_0^2$ and
$\gamma=-\Omega_0$, $\tilde{R}(t)=\mbox{e}^{\Omega_0 t}$.  (b) Corresponding contour plot. The other parameters
are $a=1$, $b=0.5$, $c=1.0$,$\Omega_0=0.1$ in Eq.~(\ref{bright3_1}).}
\label{fig6}
\end{center}
\end{figure} 
 The solitary wave solution occurs due to the balance between the negative gain term and the exponential growth of the atomic scattering length. It may be noted that, here, the overall amplitude remains constant ($A^+ = \frac{N a_B}{a_{\perp}}$). Fig.~\ref{fig6}(a) shows the bright solitary wave~(\ref{bright3_1})  for $\gamma = -\Omega_0$ and $R=\exp[2 \Omega_0 t]$. However, the width of the wave ($\propto\frac{1}{\sqrt{\tilde{R}(t)}}=\exp(-\frac{1}{2}\Omega_0 t)$) decreases exponentially with time as seen from Fig.~\ref{fig6}(b).

For the safe range parameter  $\gamma= -\Omega_0 $ the corresponding dark soliton solution of (\ref{1dGPE}) is given by the expression
\begin{align}
\vert \Phi^-(z,t) \vert = &  \frac{N a_B}{2 \beta a_{\perp}} \left\{c^2+4 \beta^2 \tanh^2\left[ \left(\frac{N a_B}{a_{\perp}}\exp(\Omega_0 t) \right)z -\left(\frac{N^3 a_B^3   }{ 2 \beta^2 a_{\perp}^3}b +\frac{N^2 a_B^2 }{2 \beta a_{\perp}^2}c\right) \frac{\exp(2\Omega_0 t)}{2 \Omega_0}\right] \right\}^{\frac{1}{2}}. \label{dark3_1}
\end{align}
One may again note that the amplitude remains constant ($A^- = \frac{N a_B}{a_{\perp}}$) in the above expression.
\begin{figure}[!ht] 
\begin{center}
\includegraphics[width=0.85\linewidth]{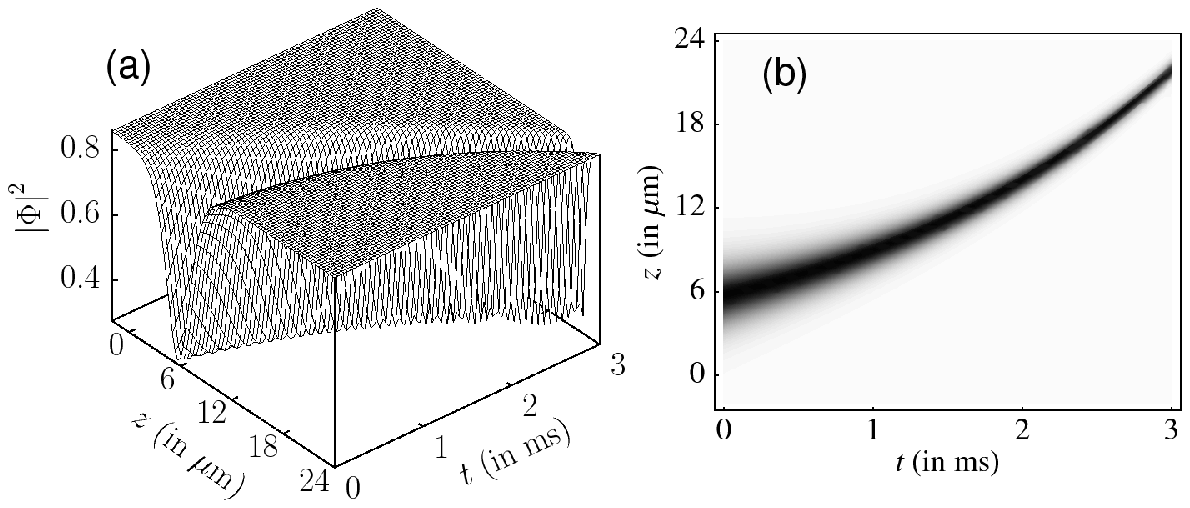}
\caption{(a) Dark soliton for $\Omega^2(t)=-\Omega_0^2$ and
$\gamma=-\Omega_0$, $\tilde{R}(t)=\mbox{e}^{\Omega_0 t}$. (b) Corresponding contour plot. The other parameters are $\beta=1/\sqrt{2}$, $b=0.5$, $c=1.0$, $\Omega_0=0.1$ in Eq.~(\ref{dark3_1}).}
\label{fig7}
\end{center}
\end{figure} 
Figs.~\ref{fig7} illustrates the dynamics of the dark solitary wave. Once again, as in the case of bright soliton, the width of the wave is decreasing exponentially with time [see Fig.~\ref{fig7}(b)], while the amplitude remains constant.

\subsubsection{$\gamma= \Omega_0 $, growth of bright/dark soliton:}
\label{case5.1.2}
For positive gain term $\gamma = \Omega_0 > 0$ case, one finds that $R$ is independent of time, which indicates that the atomic scattering length is constant. In this case, the bright soliton solution of (\ref{1dGPE}) can be written as
\begin{figure}[!ht] 
\begin{center}
\includegraphics[width=0.85\linewidth]{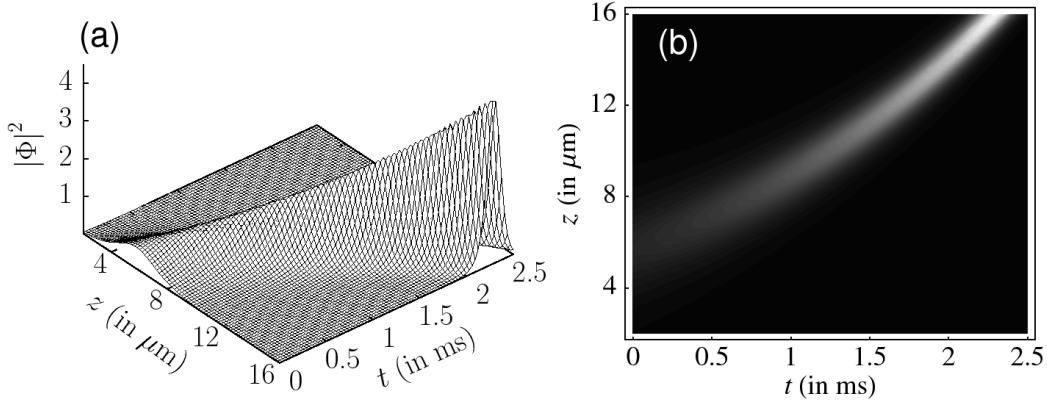}
\caption{(a) Bright soliton for $\Omega^2(t)=-\Omega_0^2$ and $\gamma=\Omega_0$, $\tilde{R}(t)=\exp (\Omega_0 t)$.  (b) Corresponding contour plot. The other parameters are $a=1$, $b=0.5$, $c=1.0$, $\Omega_0=0.1$ in Eq.~(\ref{bright3_2}).}
\label{fig8}
\end{center}
\end{figure} 
\begin{align}
\vert \Phi^+(z,t) \vert=  &\,\frac{N a_B}{a_{\perp}}  \sech\left[\left(\frac{N a_B}{a_{\perp}}\exp(\Omega_0 t) \right)z -\left(\frac{N^3 a_B^3  }{a^2 a_{\perp}^3} b+ \frac{N^2 a_B^2  }{\sqrt{2} a a_{\perp}^2}c\right) \frac{\exp(2\Omega_0 t)}{2 \Omega_0}\right] \exp\left[\Omega_0 t \right]. \label{bright3_2}
\end{align}
One may note that, in the above expression, the amplitude of bright soliton grows exponentially ($A^+ = \frac{N a_B}{a_{\perp}} \exp(\Omega_0 t)$) due to the presence of the gain term (parameters in the out of safe region), while the width ($\propto \exp(-\frac{1}{2}\Omega_0 t)$) decreases exponentially with time.
Fig.~\ref{fig8}(a) depicts the exponential growth of bright solitary wave for $\gamma =\Omega_0$ and $R =$ constant. Here the width of the wave decreases with respect to time exponentially as seen in Fig.~\ref{fig8}(b).

In this case the dark soliton solution of (\ref{1dGPE}) can be written as
\begin{align}
\vert \Phi^-(z,t) \vert = &  \frac{N a_B}{2 \beta a_{\perp}} \left\{c^2+4 \beta^2 \tanh^2\left[ \left(\frac{N a_B}{a_{\perp}}\exp(\Omega_0 t) \right)z -\left(\frac{N^3 a_B^3   }{ 2 \beta^2 a_{\perp}^3}b +\frac{N^2 a_B^2 }{2 \beta a_{\perp}^2}c\right) \frac{\exp(2\Omega_0 t)}{2 \Omega_0}\right] \right\}^{\frac{1}{2}} \exp\left[\Omega_0 t\right]. \label{dark3_2}
\end{align}
\begin{figure}[!ht] 
\begin{center}
\includegraphics[width=0.85\linewidth]{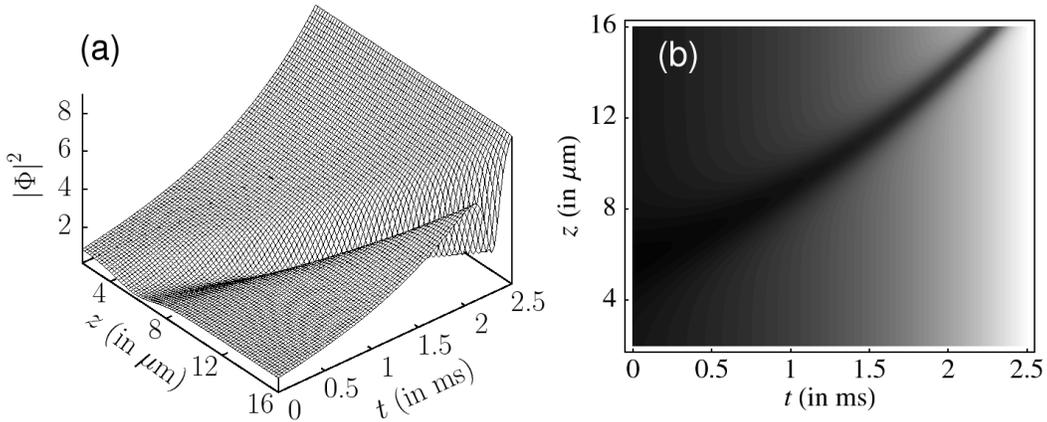}
\caption{(a) Dark soliton for $\Omega^2(t)=-\Omega_0^2$ and $\gamma=\Omega_0$, $\tilde{R}(t)=\exp (\Omega_0 t)$.  (b) corresponding contour plot. The other parameters are $\beta=1/\sqrt{2}$, $b=0.5$, $c=1.0$, $\Omega_0=0.1$ in Eq.~(\ref{dark3_2}).}
\label{fig9}
\end{center}
\end{figure} 
The exponential growth of the dark solitary wave due to the presence of gain is illustrated in Figs.~\ref{fig9}(a) and \ref{fig9}(b), where the width decreases exponentially in time.

\subsection{Case 2: $\tilde{R}=\exp(-\Omega_0 t)$}
 Next, in the case of $\tilde{R}=\exp(-\Omega_0 t)$ similar type of stable solitary waves and growth/decay of solitary waves can be identified for suitable forms of $R(t)$ and $\gamma(t)$. Also one may note here that the width of the wave will be increasing exponentially with time (i.e, proportional to $\exp(\frac{1}{2}\Omega_0 t)$) in contrast to the case discussed in Sec.~\ref{+omega}. Since the details are similar we do not present them here separately.

\subsection{Case 3: $\tilde{R}= \sech (\Omega_0 t+ \delta)$}

Next, we analyze the solutions of Eq.~(\ref{1dGPE}) for the case $\tilde{R} = \sech (\Omega_0 t+ \delta)$ and with different forms of gain term $\gamma$.

\begin{figure}[!ht] 
\begin{center}
\includegraphics[width=0.85\linewidth]{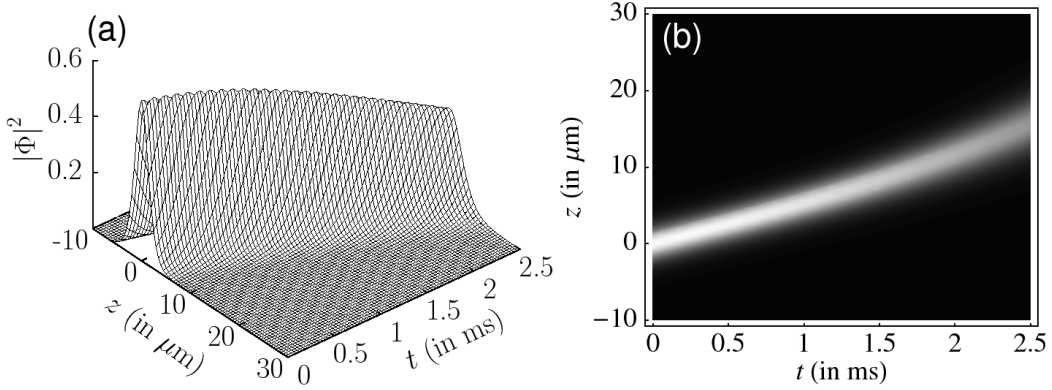}
\caption{Bright soliton : for $\Omega^2(t)=-\Omega_0^2$ and ${\gamma}=0$, $\tilde{R}(t)=\sech(\Omega_0 t+\delta)$ case. (b) The corresponding contour plot. The other parameters are $a=1$, $b=0.5$, $c=1.0$, $\Omega_0=0.1$ in Eq.~(\ref{bright_61}). }
\label{figure14}
\end{center}
\end{figure} 
\subsubsection{$\gamma=0$, decay of bright/dark soliton and two-soliton bound state:}
First we choose $\gamma = 0$. Consequently, $R$ takes the form $R=\sech (\Omega_0 t+ \delta)$ as deduced from Eqs.~(\ref{rtilde}) and (\ref{eq:rtilde_1}). The bright soliton solution of (\ref{1dGPE}) for $\gamma = 0$ can be written as

\begin{align}
\vert \Phi^+(z,t) \vert=  &\,\frac{N a_B}{a_{\perp}} \sqrt{ \sech(\Omega_0 t+\delta)}\notag \\ &\,\times \sech\left[\left(\frac{N a_B}{a_{\perp}}  \sech(\Omega_0 t +\delta)\right)z -\left(\frac{N^3 a_B^3  }{a^2 a_{\perp}^3} b+ \frac{N^2 a_B^2  }{\sqrt{2} a a_{\perp}^2}c\right)\frac{\tanh(\Omega_0 t+\delta)}{\Omega_0}\right].\label{bright_61}
\end{align}
Figs.~\ref{figure14} shows the decaying nature of the bright solitary wave 
\begin{figure}[!ht]
\begin{center}
\includegraphics[width=0.85\linewidth]{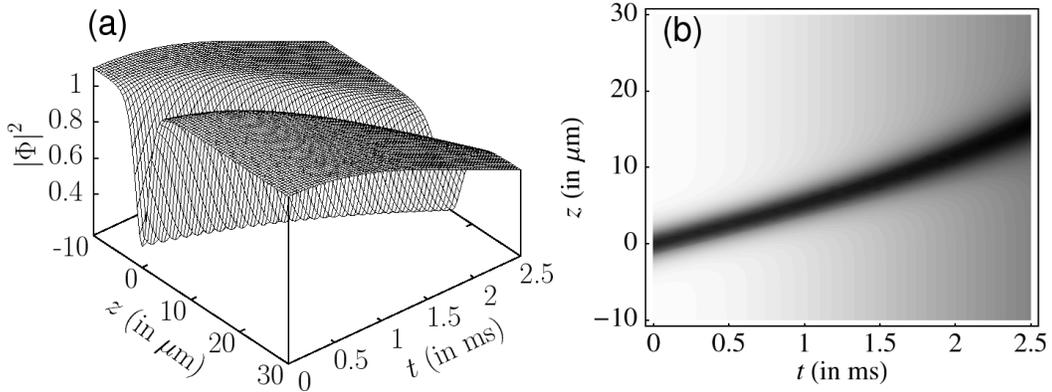}
\caption{(a) Dark soliton : for $\Omega^2(t)=-\Omega_0^2$ and $\gamma=0$, $\tilde{R(t)}=\sech(\Omega_0 t+\delta)$ case. (b) The corresponding contour plot. The other parameters are $\beta=1/\sqrt{2}$, $b=0.5$, $c=1.0$, $\Omega_0=0.1$ in Eq.~(\ref{dark_61}). }
\label{figure15}
\end{center}
\end{figure} 

\begin{figure}[!ht] 
\begin{center}
\includegraphics[width=0.85\linewidth]{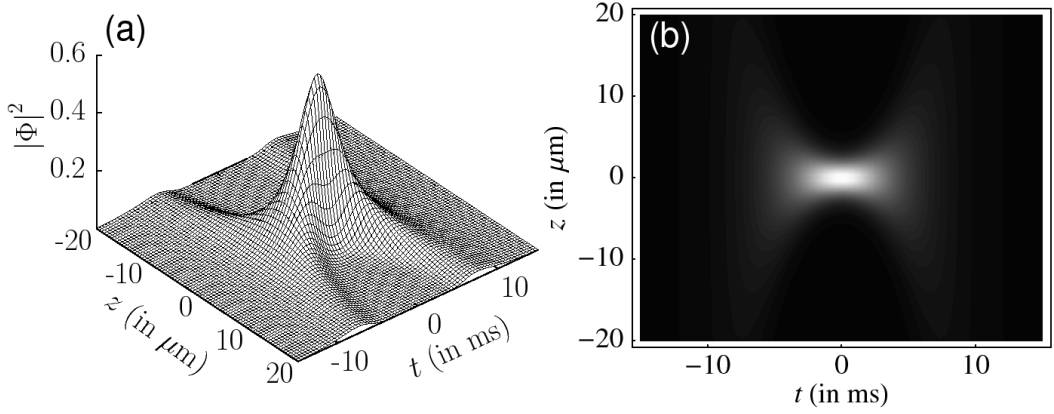}
\caption{(a) Bright soliton bound state: for $\Omega^2(t)=-\Omega_0^2$ and $\gamma=0$, $\tilde{R}(t)=\sech(\Omega_0 t+\delta)$ case. (b) The corresponding contour plot. The other parameters are $a=1$ ,$b=0$, $c=0$,$\Omega_0=0.1$ in Eq.~(\ref{bright_61}).}
\label{figure22}
\end{center}
\end{figure}

In the present case, another interesting structure similar to a bright two-soliton bound state [shown in Fig.~\ref{figure22}] occurs when the propagation term is absent in the above soliton solution (i.e, velocity is zero) when the parameters $b$ and $c$ in Eq.~(\ref{bright_61}) are chosen as $b=0$, $c=0$ . One may note that a similar kind of two-soliton like bound state has been recently identified by Atre, Panigrahi and Agarwal using a different approach~\cite{Atre2006}.

\begin{figure}[!ht]
\begin{center}
\includegraphics[width=0.85\linewidth]{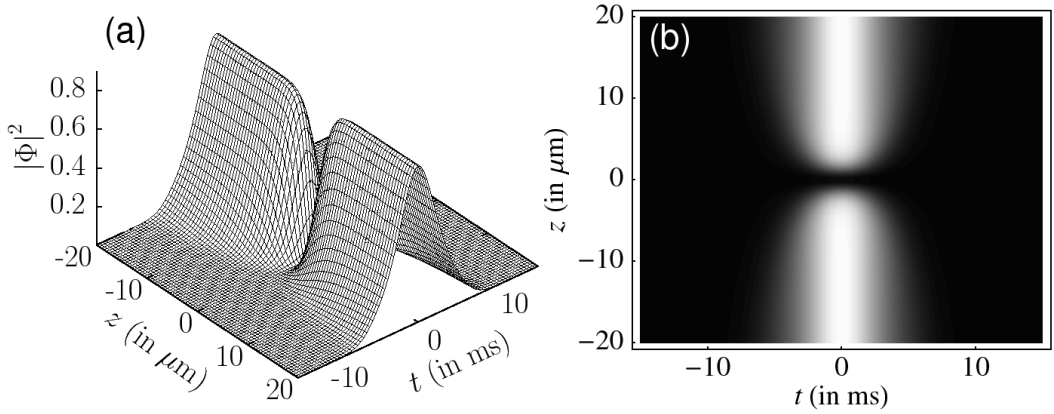}
\caption{(a) Dark soliton bound state: for $\Omega^2(t)=-\Omega_0^2$ and $\gamma=0$, $\tilde{R}(t)=\sech(\Omega_0 t+\delta)$ case. (b) The corresponding contour plot. The other parameters are $\beta=1/\sqrt{2}$, $b=0$, $c=0$,$\Omega_0=0.1$ in Eq.~(\ref{dark_61}).}
\label{figure23}
\end{center}
\end{figure}

Next, the dark soliton solution of (\ref{1dGPE}) for the present case can be written as
\begin{align}
\vert \Phi^-(z,t) \vert = &  \frac{N a_B}{2 \beta a_{\perp}}\sqrt{ \sech(\Omega_0 t+\delta)} \notag \\ &\, \times \left\{c^2+4 \beta^2 \tanh^2\left[\left(\frac{N a_B}{a_{\perp}}  \sech(\Omega_0 t +\delta)\right)z -\left(\frac{N^3 a_B^3   }{ 2 \beta^2 a_{\perp}^3}b +\frac{N^2 a_B^2 }{2 \beta a_{\perp}^2}c\right)\frac{\tanh(\Omega_0 t+\delta)}{\Omega_0} \right]\right\}^{\frac{1}{2}}. \label{dark_61}
\end{align}
Figs.~\ref{figure15} shows the decaying nature of the dark solitary wave solution for $b \neq 0$ and $c \neq 0$, where the width also increases exponentially in time. Similar to the case of bright soliton solution, a dark two-soliton like bound state can arise when $b=0$, $c=0$ in Eq.~(\ref{dark_61}).  Figs.~\ref{figure23} shows such a dark two-soliton like bound state for $b=0$, $c=0$ in Eq.~(\ref{dark_61}).

\begin{figure}[!ht]
\begin{center}
\includegraphics[width=0.85\linewidth]{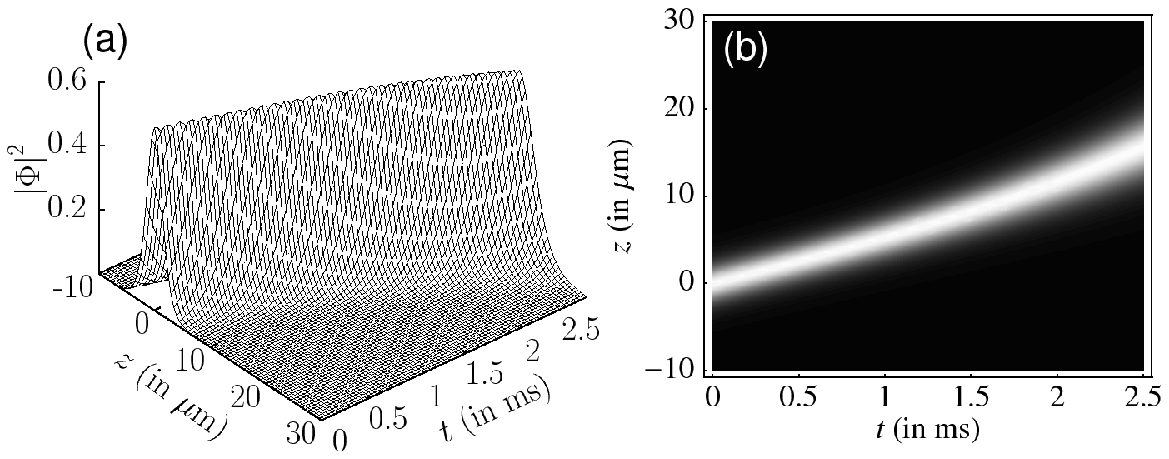}
\caption{(a) Bright soliton: for $\Omega^2(t)=-\Omega_0^2$ and $\gamma=\Omega_0 \tanh( \Omega_0 t +\delta)$, $\tilde{R}(t)=\sech(\Omega_0 t+\delta)$ case.(b) The corresponding contour plot. The other parameters are $a=1$ ,$b=0.5$, $c=1.0$,$\Omega_0=0.1$ in Eq.~(\ref{Bright_62}).}
\label{figure16}
\end{center}
\end{figure} 
\begin{figure}[!ht]
\begin{center}
\includegraphics[width=0.85\linewidth]{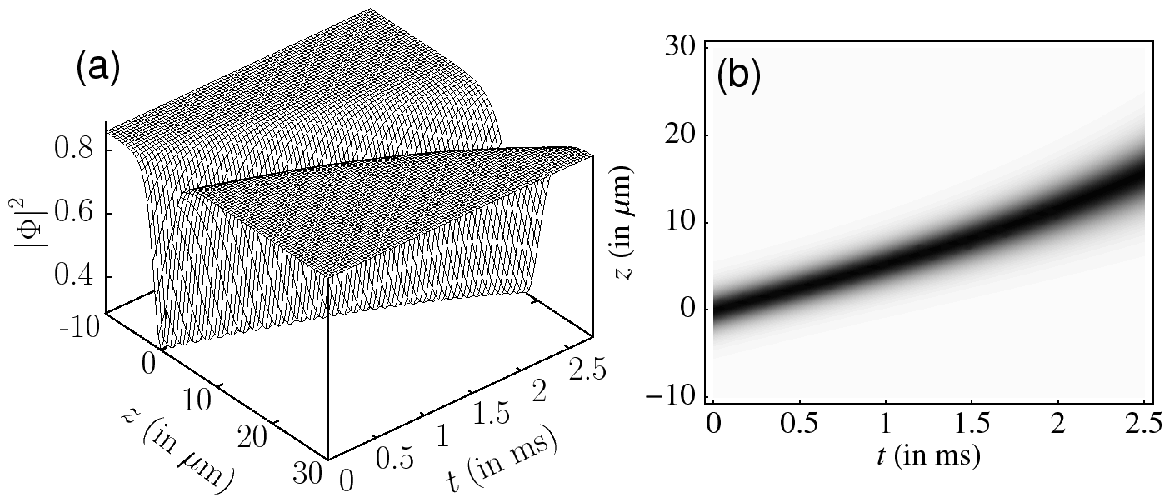}
\caption{(a) Dark soliton: for $\Omega^2(t)=-\Omega_0^2$ and $\gamma=\Omega_0 \tanh( \Omega_0 t +\delta)$, $\tilde{R}(t)=\sech(\Omega_0 t+\delta)$ case.(b) The corresponding contour plot. The other parameters are $\beta=1/\sqrt{2}$, $b=0.5$, $c=1.0$,$\Omega_0=0.1$ in Eq.~(\ref{dark_62}).}
\label{figure17}
\end{center}
\end{figure} 

\subsubsection{$\gamma=\Omega_0 \tanh\left[\Omega_0 t+ \delta \right]$, bright/dark soliton:}
Next let us choose the form of the gain term as $\gamma = \Omega_0 \tanh\left[\Omega_0 t+ \delta \right]$.  Then the form of the atomic scattering length becomes $R = \sech^2 (\Omega_0 t+ \delta)$, as obtained from Eqs.~(\ref{rtilde}) and (\ref{eq:rtilde_1}). Consequently, the bright soliton solution of (\ref{1dGPE}) can be written as
\begin{align}
\vert \Phi^+(z,t) \vert=  &\,\frac{N a_B}{a_{\perp}}  \sech\left[\left(\frac{N a_B}{a_{\perp}}  \sech(\Omega_0 t +\delta)\right)z -\left(\frac{N^3 a_B^3  }{a^2 a_{\perp}^3} b+ \frac{N^2 a_B^2  }{\sqrt{2} a a_{\perp}^2}c\right)\frac{\tanh(\Omega_0 t+\delta)}{\Omega_0}\right], \label{Bright_62}
\end{align}

Fig.~\ref{figure16}(a) shows the corresponding bright solitary wave. One may note that the amplitude of the wave is constant ($A^+=\frac{N a_B}{a_{\perp}}$) over all time due to the balance between the gain and atomic scattering length but the width of  the wave increases with respect to time (proportional to $\sqrt{\cosh(\Omega_0 t+\delta)}$) as shown in Fig.~\ref{figure16}(b).

 Next, the corresponding dark soliton solution of (\ref{1dGPE}) can be written as
\begin{align}
\vert \Phi^-(z,t) \vert = &  \frac{N a_B}{2 \beta a_{\perp}}\left\{c^2+4 \beta^2 \tanh^2\left[ \left(\frac{N a_B}{a_{\perp}}  \sech(\Omega_0 t +\delta)\right)z -\left(\frac{N^3 a_B^3   }{ 2 \beta^2 a_{\perp}^3}b +\frac{N^2 a_B^2 }{2 \beta a_{\perp}^2}c\right)\frac{\tanh(\Omega_0 t+\delta)}{\Omega_0} \right]\right\}^{\frac{1}{2}} \label{dark_62}
\end{align}
which is shown in Figs.~\ref{figure17}. Here the amplitude, $A^-=\frac{N a_B}{a_{\perp}}$, remains constant while the width increases with time.

\subsubsection{$\gamma = a_1 t-a_2 \sin(\kappa t)$, collapse and revival of the bright/dark solitary waves:}
Finally, we consider the case with periodically varying gain term of the form  $\gamma =a_1 t-a_2 \sin(\kappa t)$. In this case, one may obtain the expressions for the atomic scattering length  from Eqs.~(\ref{rtilde}) and (\ref{eq:rtilde_1}) as
\begin{align}
R = \exp\left[-a_1 \frac{t^2}{2}-\frac{\cos(\kappa t)}{ \kappa}  \right] \sech(\Omega_0 t+ \delta).\end{align}
Here again we observe the periodic collapse and revival of the bright and dark solitary waves, similar to the case of the confining potential as shown in Fig.~\ref{figure2}. However, the amplitude is decaying with time ($\sim \sqrt{ \sech(\Omega_0 t+\delta)}$ $\times\exp\left[a_1 \frac{t^2}{4}+\frac{\cos(\kappa t)}{2 \kappa}  \right]$) while in Fig.~\ref{figure2} it increases with time ($\sim\sqrt{ \sec(\Omega_0 t+\delta)} $ $\times\exp\left[a_1 \frac{t^2}{4}+\frac{\cos(\kappa t)}{2 \kappa}  \right]$). However, we do not present the further details as the qualitative features are similar to the case~\ref{case3_fig2}.

In the above we have identified various interesting soliton solutions that exist in the case of time-independent harmonic potentials. A summary of these results are tabulated in Table~\ref{table3} for easy reference. It is also of wide  interest at present to look at the nature of the solutions in the presence of time varying harmonic trap potentials as experiments in BEC with time varying harmonic trap potentials are also possible. In the following section, we consider certain interesting soliton solutions that exist in a select set of time-dependent traps.
\begin{table}[!ht]
\centering
\caption{Nature of soliton solutions for different forms of trap potential $\Omega^2(t)$, modulated scattering length $R(t)$  and gain/loss term$\gamma(t)$ in the time-independent harmonic traps}
\label{table3}
\begin{tabular}{|l|ll|c|l|}
\hline  \multicolumn{1}{|c|}{$\Omega^2(t)$} & \multicolumn{2}{c|}{$ R(t)$} & \multicolumn{1}{c}{$\gamma(t)$} & \multicolumn{1}{|c|}{Nature of Solution} \\
\hline
   & (i)&$\sec(\Omega_0 t)$ & $0$& periodic peaks in amplitude \\ & &&&\\

   $\Omega_0^2$&(ii)&$\sec(\Omega_0 t) \displaystyle\exp\left[-\frac{a_1}{2} t^2- a_2 \frac{\cos(\kappa t)}{\kappa}\right]$ & $a_1 t-a_2 \sin(\kappa t)$ & periodic collapse \& revival with amplitude growth\\ &&& & \\
  &(iii)&$\sec^2(\Omega_0 t) $ & $-\Omega_0 \tan(\Omega_0 t )$ &  snake like  structure with constant amplitude \\ & & && \\
\hline  &&&&\\ & (i)&$\exp(2 \Omega_0 t)$ &
 $-\Omega_0$& amplitude constant/width decreases\\&&&&\\
  & (ii)&constant &
 $\Omega_0$& amplitude grows/width decreases\\&&&&\\
  $-\Omega_0^2$ &(iii)&$\exp(-2 \Omega_0 t)$ &
 $\Omega_0$& amplitude constant/width increases\\&&&&\\
   & (iv)&constant &
 $-\Omega_0$& amplitude decreases/width increases\\&&&&\\
 &(v)&$\sech(\Omega_0 t)$&$0$ & amplitude decreases/width increases\\&&&&or bound state occurs\\&&&&\\
 &(vi)&$\sech^2(\Omega_0 t)$&$\Omega_0 \tanh(\Omega_0 t)$ & amplitude constant/width increases\\&&&&\\
 &(vii)&$\sech(\Omega_0 t) \exp(\Omega_0 t)$&$-\frac{\Omega_0}{2} $ & amplitude decreases/width increases\\&&&&\\
 &(viii)&$\sech(\Omega_0 t) \exp\left[-\frac{a_1}{2} t^2- a_2 \frac{cos(\kappa t)}{\kappa}\right]$ & $a_1 t-a_2 \sin(\kappa t)$ & periodic collapse \& revival with decaying amplitude\\
\hline
\end{tabular}
\end{table}

\section{Bright and dark solitons in the time-dependent harmonic traps}
\label{de}

In this section, we consider certain interesting time-dependent trap potentials for which Eq.~(\ref{1dGPE}) exhibits soliton solutions. We again identify a variety of soliton solutions by suitably choosing the gain and atomic scattering length.

\subsection{Double modulated periodic potential}

First let us choose the double modulated periodic potential, $\Omega^2(t)$, of the form
\begin{align}
\Omega^2(t)=-\Omega_0\left(\frac{\Omega_0}{2}\left[1-\cos(2\lambda t)\right]- \lambda \cos(\lambda t)\right). \label{double_period}
\end{align}
Substituting the above form of $\Omega^2(t)$ into Eq.~(\ref{con:eqn}), one may obtain a particular solution for $\tilde{R}$  as [see Table~\ref{table1}]
\begin{align}
\tilde{R}=\exp\left[\frac{-\Omega_0}{\lambda}\cos(\lambda t)\right].
\end{align}

\subsubsection{ $\gamma = -\Omega_0  \sin(\lambda t)$, periodic collapse and revival of the two-soliton bound state}

In this case, the soliton solution can be identified by choosing a periodically varying gain or loss term of the form $\gamma = -\Omega_0  \sin(\lambda t)$. Then from  Eqs.~(\ref{rtilde}) and (\ref{eq:rtilde_1}) one can get the atomic scattering length as
\begin{align}
R = \exp\left[\frac{-2 \Omega_0 t}{\lambda} \cos(\lambda t)\right].
\end{align}
In this case an interesting type of periodic bound state bright soliton solution of (\ref{1dGPE}) can be obtained (for $b=0$, $c=0$) as
\begin{align}
\vert \Phi^+(z,t) \vert=  &\,\frac{N a_B}{a_{\perp}} \exp \left[\frac{-\Omega_0\cos(\lambda t)}{\lambda}\right]\notag \\ &\, \times \sech\left[ \left(\frac{N a_B}{a_{\perp}} \exp\left[\frac{ -\Omega_0\cos(\lambda t)}{\lambda}\right]\right)z \right]. \label{Bright_85}
\end{align}
\begin{figure}[!ht]
\begin{center}
\includegraphics[width=0.85\linewidth]{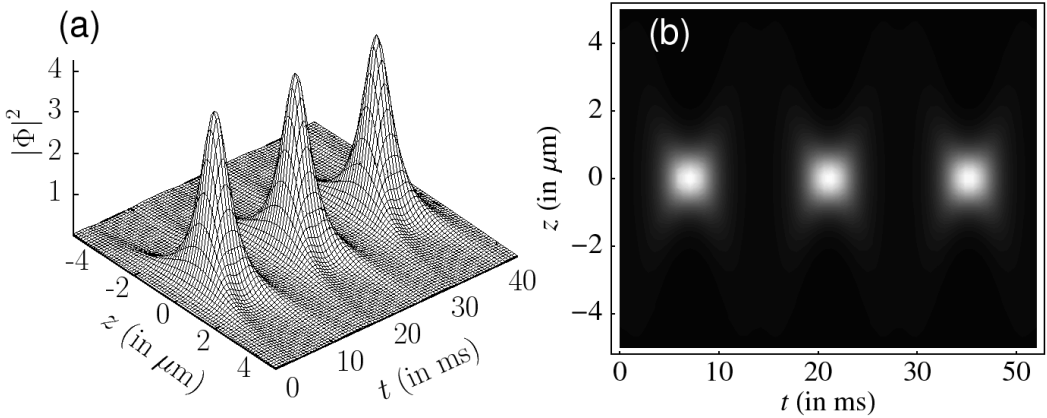}
\caption{(a) Bright soliton bound state: for $\Omega^2(t)=-\Omega_0 (\Omega_0/2 (1-\cos(2 \lambda t))-\lambda \cos(\lambda t))$ and $\gamma=-\Omega_0 \sin(\lambda t)$, $\tilde{R}(t)=\exp[\frac{-\Omega_0}{\lambda} \cos(\lambda t)]$ case. (b) The corresponding contour plot.  The parameters are $a=1$, $b=0$, $c=0$, $\Omega_0=\lambda=0.1$ in Eq.~(\ref{Bright_85}).}
\label{figure24}
\end{center}
\end{figure}%

As before, one may observe that the amplitude of the soliton is $A^+=\frac{N a_B}{a_{\perp}} \exp \left[\frac{-\Omega_0\cos(\lambda t)}{\lambda}\right]$ and width is proportional to $\exp\left[\frac{ \Omega_0\cos(\lambda t)}{2\lambda}\right]$ indicating periodic collapse and revival of the wave function.
In the above, we see an additional feature in the form of a periodic collapse and revival of the bright two-soliton bound state of the matter wave for $b = 0$ and $c = 0$. The bound state indicates the absence of propagation term (for $b = 0$ and $c = 0$). However collapse and revival phenomena occur due to the fact that amplitude ($\sim \exp \left[\frac{-\Omega_0\cos(\lambda t)}{\lambda}\right]$) is periodically varying with time. This is illustrated in Fig.~\ref{figure24}. 

\begin{figure}[!ht]
\begin{center}
\includegraphics[width=0.85\linewidth]{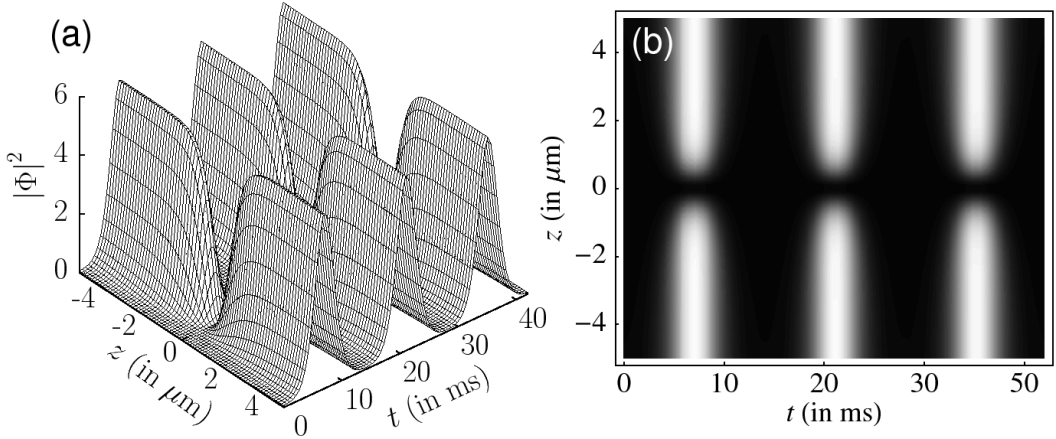}
\caption{(a) Dark soliton: for $\Omega^2(t)=-\Omega_0 (\Omega_0/2 (1-\cos(2 \lambda t))-\lambda \cos(\lambda t))$ and $\gamma=-\Omega_0 \sin(\lambda t)$, $\tilde{R}(t)=\exp[\frac{-\Omega_0}{\lambda} \cos(\lambda t)]$ case. (b) The corresponding contour plot.  The parameters are $\beta=1/\sqrt{2}$, $b=0$, $c=0$, $\Omega_0=\lambda=0.1$ in Eq.~(\ref{dark_85}).}
\label{figure25}
\end{center}
\end{figure}
The corresponding periodic bound state dark soliton solution of (\ref{1dGPE}) is given as
\begin{align}
 \vert&\Phi^-(z,t) \vert =   \frac{N a_B}{2 \beta a_{\perp}}  \exp \left[\frac{-\Omega_0\cos(\lambda t)}{\lambda}\right]  \notag \\ &\times \left\{c^2+4 \beta^2 \tanh^2\left[ \left(\frac{N a_B}{a_{\perp}} \exp\left[\frac{ -\Omega_0\cos(\lambda t)}{\lambda}\right]\right)z\right]\right\}^{\frac{1}{2}}. \label{dark_85}
\end{align}
 Fig.~\ref{figure25} shows the collapse and revival of the dark two-soliton bound state.

\subsection{Step-wise time-dependent trap potential}

Next, let us consider a step-function modulation potential, $\Omega(t)$ of the form
\begin{align}
\Omega^2(t) = -\frac{\Omega_0^2}{2}\left[1-\tanh\left(\frac{\Omega_0}{2} t\right)\right]. \label{step_poten}
\end{align}
Using the above form of $\Omega(t)$ in the integrability condition (\ref{con:eqn}), we get a particular solution for $\tilde{R}$, see Table~\ref{table1},
\begin{align}
\tilde{R}=1+\tanh\left(\frac{\Omega_0}{2} t\right).
\end{align}

\subsubsection{Case 1: $\gamma(t) =-\frac{\Omega_0}{2 }\left[1 - \tanh\left(\frac{\Omega_0}{2} t\right)\right]$, bright/dark soliton}
If we choose the time-dependent gain or loss term as
\begin{align}
\gamma(t) =-\frac{\Omega_0}{2 }\left[1 - \tanh\left(\frac{\Omega_0}{2} t\right)\right] \label{step_gain},
\end{align}
then from  Eqs.~(\ref{rtilde}) and (\ref{eq:rtilde_1}) the form of the atomic scattering length $R$ turns out to be
\begin{align}
R=\frac{4 \exp(2 \Omega_0 t)}{\left[1+\exp( \Omega_0 t)\right]^2}.
\end{align}
Consequently, the bright soliton solution of (\ref{1dGPE}) for the time-dependent potential (\ref{step_poten}) with the above loss term (\ref{step_gain}) can be written as
\begin{align}
\vert \Phi^+(z,t) \vert= & \, \frac{N a_B}{a_{\perp}}  \sech\Biggr[ \frac{N a_B}{a_{\perp}} z\left[1+\tanh\left(\frac{\Omega_0 t}{2}\right)\right] -\left(\frac{N^3 a_B^3  }{a^2 a_{\perp}^3} \frac{b}{\Omega_0}+ \frac{N^2 a_B^2  }{\sqrt{2}  a a_{\perp}^2} \frac{c}{\Omega_0}\right) \notag \\
& \, \times  \left\{2 \Omega_0 t+4 \log \left[\cosh \left(\frac{\Omega_0 t}{2}\right)\right]-2 \tanh\left(\frac{\Omega_0 t}{2}\right)\right\} \Biggr].   \label{Bright_86}
\end{align}
\begin{figure}[!ht] 
\begin{center}
\includegraphics[width=0.85\linewidth]{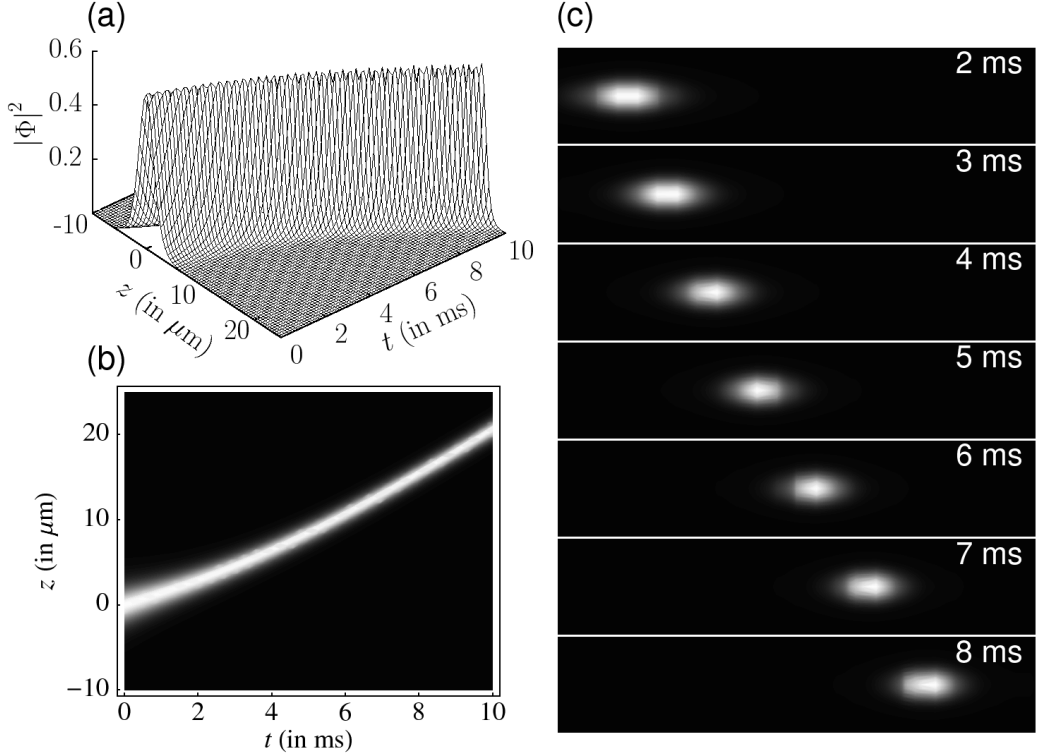}
\caption{(a) Bright soliton: for $\Omega^2(t)=-\Omega_0^2 \left[1 - \tanh(\frac{\Omega_0}{2} t)\right]$, $\tilde{R}(t)=1 + \tanh(\frac{\Omega_0}{2} t)$ and $\gamma=-\frac{\Omega_0}{2 }[1 - \tanh(\frac{\Omega_0}{2} t)]$ case. (b) The corresponding contour plot in the $t-x$ plane and (c) Snapshots of soliton profile in the $z-x$ plane at different instances of time. The parameters are $a=1.0$, $b=0.2$, $c=0.2$, $\Omega_0=0.1$ in Eq.~(\ref{Bright_86}).}
\label{figure26}
\end{center}
\end{figure} 
The bright solitary wave solution, in this case, is shown in Figs.~\ref{figure26}(a) and \ref{figure26}(b). One may note that the amplitude of the wave remains constant ($A^+=\frac{N a_B}{a_{\perp}}$) while the width gradually decreases ($\sim 1/\sqrt{\left[1+\tanh\left(\frac{\Omega_0 t}{2}\right)\right]}$) with time. 

It is also of interest to look at the solution of the original GP equation (\ref{3dGPE}) with the trap potential (\ref{step_poten}) with time-dependent loss as in (\ref{step_gain}). Fig.~\ref{figure26}(c) shows snapshots of the bright soliton solution in the $z-x$ plane at different instances of time. One may note that the set of soliton profiles which are shown Fig.~\ref{figure26}(c) resembles the one that is obtained in certain recent experiments \cite{Khaykovich2002,Strecker2002}. However, these experiments were carried out for the formation of bright matter wave solitons in ultracold $^{7}$Li gas, with confining/expulsive time independent harmonic trap potentials. Here we have identified a similar kind of soliton solution in the step-wise modulated harmonic potential, which is experimentally feasible to verify in the future. In Ref~\cite{Xue2005}, Xue studied the bright soliton solution without gain term  for this case. 

\begin{figure}[!ht] 
\begin{center}
\includegraphics[width=0.85\linewidth]{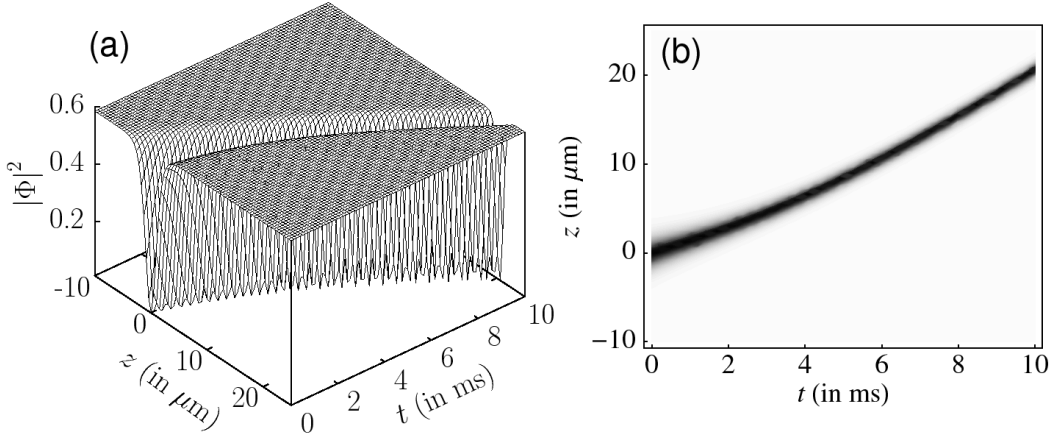}
\caption{ (a) Dark soliton: for $\Omega^2(t)=-\Omega_0^2 \left[1 - \tanh(\frac{\Omega_0}{2} t)\right]$, $\tilde{R}(t)=1 + \tanh(\frac{\Omega_0}{2} t)$ and $\gamma=-\frac{\Omega_0}{2 }[1 - \tanh(\frac{\Omega_0}{2} t)]$ case. (b) The corresponding contour plot. The parameters are $\beta=1/\sqrt{2}$, $b=0.2$, $c=0.2$, $\Omega_0=0.1$in Eq.~(\ref{dark_86}). }
\label{figure27}
\end{center}
\end{figure}

Next, in this case the dark soliton solution can be written as
\begin{align}\label{dark_86}
 \vert\Phi^-(z,t) \vert = &   \frac{N a_B}{2 \beta a_{\perp}}\left\{c^2+4 \beta^2 \tanh^2\left[ \frac{N a_B}{  a_{\perp}} z\left[1+\tanh\left(\frac{\Omega_0 t}{2}\right)\right] \right. \right.\notag \\ & \left. \left.-\left(\frac{N^3 a_B^3   }{ 2 \beta^2 a_{\perp}^3}\frac{b}{\Omega_0} +\frac{N^2 a_B^2 }{2 \beta a_{\perp}^2}\frac{c}{\Omega_0}\right) \left\{2 \Omega_0 t+4 \log \left[\cosh \left(\frac{\Omega_0 t}{2}\right)\right]-2 \tanh\left(\frac{\Omega_0 t}{2}\right)\right\}\right]\right\}^{\frac{1}{2}}
\end{align}
The dark soliton solution is shown in Fig. \ref{figure27}. In this case also the amplitude of the wave remains constant ($\frac{N a_B}{a_{\perp}}$), while the width gradually decreases ($\sim 1/\sqrt{\left[1+\tanh\left(\frac{\Omega_0 t}{2}\right)\right]}$) with time.

\subsubsection{Case 2: $\gamma=\frac{\Omega_0}{2 } > 0$, growth of bright/dark soliton}

Finally, we consider a positive gain term of the form $\gamma=\frac{\Omega_0}{2 } > 0$. For this choice of $\gamma$ Eqs.~(\ref{rtilde}) and (\ref{eq:rtilde_1}) yield the condition on the the atomic scattering length as $R=\sech \left(\frac{ \Omega_0 t}{2}\right) $. In this case we see a growth of the bright/dark solitary wave similar to the case~\ref{case5.1.2} with time independent potential. Here the amplitude $\frac{N a_B}{a_{\perp}}  \sqrt{\sech\left[\frac{\Omega_0}{2}\right]}.\exp\left[\frac{\Omega_0 t}{2}\right]$ increases in time, while the width gradually decreases ($\sim 1/\sqrt{\left[1+\tanh\left(\frac{\Omega_0 t}{2}\right)\right]}$) with time. Since the details are similar to the case~\ref{case5.1.2},  we do not present them here separately.

A summary of the results of time-dependent harmonic traps is tabulated in Table~\ref{table4} for ready reference.

\begin{table}[!ht]
\centering
\caption{Nature of soliton solutions for different forms of $\Omega^2(t)$, $R(t)$  and $\gamma(t)$ in the time-dependent harmonic traps.}
\label{table4}
\begin{tabular}{|l|l|ll|l|}
\hline  \multicolumn{1}{|c|}{$\Omega^2(t)$} & \multicolumn{1}{c|}{$ R(t)$} & \multicolumn{2}{c}{$\gamma(t)$} & \multicolumn{1}{|c|}{Nature of Solution} \\
\hline &&&&\\
$-\Omega_0\left(\frac{\Omega_0}{2}\left[1-\cos(2\lambda t)\right]- \lambda \cos(\lambda t)\right)$  &  $\exp\left[\frac{-\Omega_0}{\lambda}\cos(\lambda t)\right]$ &  & $-\Omega_0  \sin \lambda t$ &  periodic collapse \& revival  of \\ & &&& two soliton like bound state\\
\hline  &&&&\\
$-\frac{\Omega_0^2}{2}\left[1-\tanh\left(\frac{\Omega_0}{2} t\right)\right]$ & $ \displaystyle\frac{4 \exp(2 \Omega_0 t)}{\left[1+\exp( \Omega_0 t)\right]^2} $ & (i) &
$-\displaystyle\frac{\Omega_0}{2 }\left[1 - \tanh\left(\frac{\Omega_0}{2} t\right)\right]$ & bright/dark soliton amplitude constant\\&&&& while width gradually decreases\\
  &  &
 (ii) & $\displaystyle \frac{\Omega_0}{2 } > 0$ & bright/dark soliton amplitude grows \\&&&&\\
\hline
\end{tabular}
\end{table}

\section{Summary and conclusions}\label{con}

In summary, we have investigated the exact bright and dark solitary wave solutions of an effective 1D BEC by assuming that the interaction energy is much less than the kinetic energy in the transverse direction. In particular we have shown that the effective 1D equation resulting from the GP equation can be transformed into the standard soliton (bright/dark) possessing, completely integrable 1D NLS equation by effecting a change of variables of the coordinates and the wave function. We have focussed on both confining and expulsive harmonic trap potentials separately and treated the atomic scattering length, gain/loss term and trap frequency as experimental control parameters by modulating them as a function of time. Various types of bright and dark soliton solutions have been deduced in the context of effective 1D Bose-Einstein condensation for different forms of unmodulated and modulated harmonic potentials by suitably tailoring the atomic scattering legnth and atom gain/loss term. First, we have observed the periodic oscillating solitons, collapse and revival of condensate and  snake-like solitons for time independent confining harmonic potential. Next we have  shown the existence of stable solitons, soliton growth and decay and formation of two-soliton bound state for time independent expulsive harmonic potential. However when the trap frequency is also modulated, we have shown the phenomena of collapse and revival of two-soliton like bound state formation of the condensate for double modulated periodic potential and bright and dark solitons for step-wise modulated potentials. Finally we have shown the snapshots of the bright soliton solution of 3D BEC in the $z-x$ plane at different instances of time for step-wise modulated harmonic potential, which is similar to certain experimental~\cite{Khaykovich2002,Strecker2002} results for time-independent potential. It appears that suitable experiments can be devised to identify these structures in BECs for time-dependent potentials. The experiments in BEC with these temporally modulated potentials are possible in future. Finally we will separately discuss the soliton interactions and the multi-soliton solutions of BECs with time varying parameters by extending our above analysis.

\ack

This work is supported in part by Department of Science and Technology (DST),
Government of India - DST-IRHPA project (ML), and DST Ramanna Fellowship (ML).The work of PM forms part of DST-FTYS (Department of Science and Technology, Government of India) project (Ref. No. SR/FTP/PS-79/2005).

\appendix

\section{Validity for 1D approximation for all cases in this paper}\label{app:validity}
\begin{figure}[!ht]
\begin{center}
\includegraphics[width=0.75\linewidth]{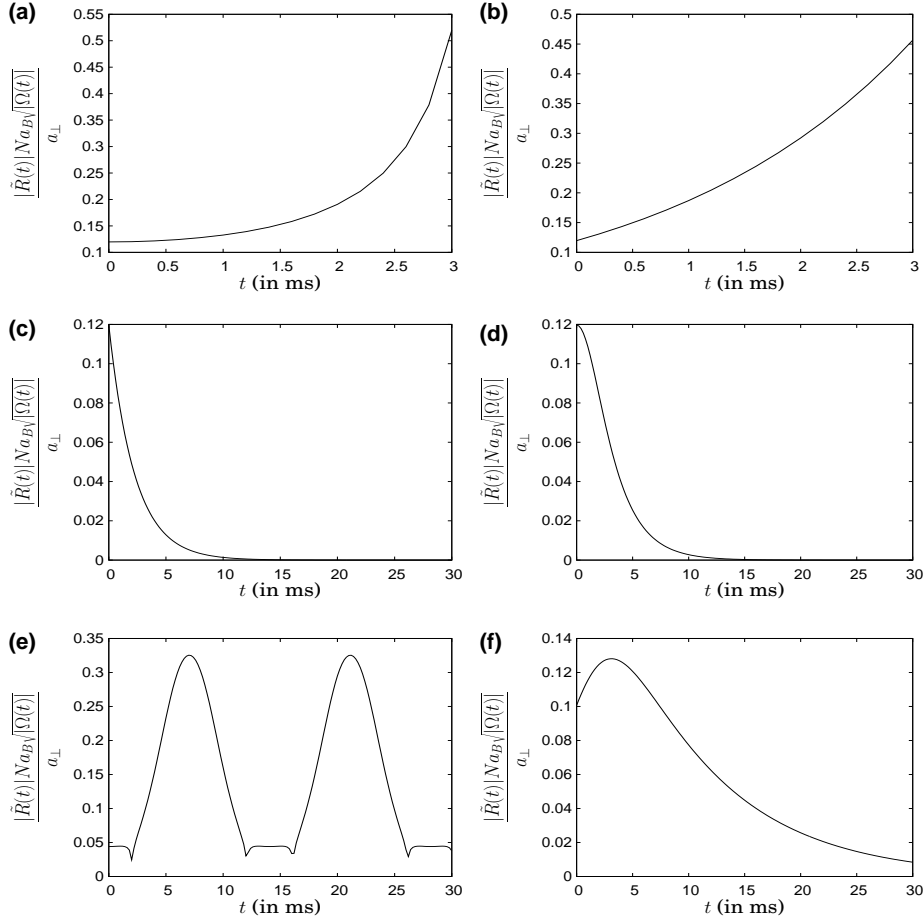}
\caption{Plot $\frac{\tilde{R}(t) N a_B \sqrt{\vert \Omega(t) \vert} }{a_{\perp}}$ vs $t$ :- (a) for $\tilde{R}(t)=\sec(\Omega_0 t +\delta)$ and $\vert \Omega(t) \vert = \Omega_0$ [see Sec. 4.1], (b) for $\tilde{R}(t)=\exp(\Omega_0 t )$ and $\vert \Omega(t) \vert=\Omega_0$ [see Sec. 5.1], (c) for  $\tilde{R}(t)=\exp(-\Omega_0 t )$ and $\vert \Omega(t) \vert=\Omega_0$ [see Sec. 5.2], (d) for $\tilde{R}(t)=\sech(\Omega_0 t +\delta)$ and $\vert \Omega(t) \vert=\Omega_0$ [see Sec. 5.3], (e) for $\tilde{R}=\exp\left[\frac{-\Omega_0}{\lambda}\cos(\lambda t)\right]$ and $\vert \Omega(t) \vert= \sqrt{\vert-\Omega_0\left(\frac{\Omega_0}{2}\left[1-\cos(2\lambda t)\right]- \lambda \cos(\lambda t)\right) \vert}$ [see Sec. 6.1] and (f) for $\tilde{R}(t)=1+\tanh\left(\frac{\Omega_0}{2} t\right)$ and $\vert \Omega(t) \vert=\sqrt{\vert -\frac{\Omega_0^2}{2}\left[1-\tanh\left(\frac{\Omega_0}{2} t\right)\right]\vert}$ [see Sec. 6.2]. The parameter values are chosen as $\Omega_0=0.1$, $\delta=0$, $\lambda=0.1$, $N=10^4$, $a_{\perp}=1.4 \mu$m. For Figs. (a-d) see also Table~1 and for (e) and (f) see Table~2 for more details. }
\label{1d_app}
\end{center}
\end{figure}

The validity of the criteria for 1D GP equation with time varying parameters $a_s(t)$, $\Gamma(t)$ and $a_z(t)$ is given by Eq.(6). Now we substitute $a_s(t)=\frac{R(t) a_B}{2}$, $\Gamma(t)= \omega_{\perp}\gamma(t)$, and $a_z(t)=a_{\perp} \sqrt{\vert\Omega(t)\vert}$ in Eq.(6) to get 
\begin{align} \label{Eq:val1} 
\frac{ R(t) N a_B \sqrt{\vert \Omega(t) \vert} }{a_{\perp}}\exp\left[\int \gamma(t) dt\right] \ll 1.
 \end{align}
Then using the form   $\tilde{R}(t)= R(t) \exp\left[\int \gamma(t) dt\right]$, see Eq.~(\ref{rtilde}), Eq.~(\ref{Eq:val1}) for the validity of the 1D approximation can be rewritten as 
 \begin{align}
 \frac{\tilde{R}(t) N a_B \sqrt{\vert \Omega(t) \vert} }{a_{\perp}} \ll 1
\end{align}
The above criterion should be satisfied for sufficiently long time for given form of $\Omega(t)$ (which is related to $a_z(t)$)  and $\tilde{R}(t)$ (which is related $a_s(t)$ and $\Gamma(t)$). The various forms of $\Omega(t)$ discussed in this paper are given in Tables ~1 and~2. Fig.~\ref{1d_app} shows the value of $\frac{\vert \tilde{R}(t)\vert N a_B \sqrt{\vert \Omega(t) \vert} }{a_{\perp}}$  with time $t$ for all forms of $\tilde{R}(t)$, $\Omega(t)$ studied in this paper. 

Fig.~\ref{1d_app}(a) shows the validity criteria of the 1D GP equation for $\tilde{R}(t)=\sec(\Omega_0 t +\delta)$ and $\Omega^2(t)=\Omega_0^2$.  The validity criteria is satisfied for several milliseconds for 
$\tilde{R}(t)=\exp(\Omega_0 t )$ and $\Omega^2(t)=-\Omega_0^2$ which is shown in Fig~\ref{1d_app}(b), which is sufficiently long for BEC experiments, See Sec.~5.1 for various 
forms of $\gamma(t)$. Figs.~\ref{1d_app}(c) and ~\ref{1d_app}(d) show  the validity criteria valid for all values of time by choosing $\tilde{R}(t)=\exp(-\Omega_0 t )$ and 
$\tilde{R}(t)=\sech(\Omega_0 t +\delta)$ respectively for $ \Omega^2(t)=-\Omega_0^2$, see Sec.~5.2 and 5.3. For Figs.~(\ref{1d_app} a)-(\ref{1d_app}d), see also Table~1.  Fig.~\ref{1d_app}(e) describes the validity criteria 
for $\tilde{R}=\exp\left[\frac{-\Omega_0}{\lambda}\cos(\lambda t)\right]$ and $\vert \Omega(t) \vert= \sqrt{\vert-\Omega_0\left(\frac{\Omega_0}{2}\left[1-\cos(2\lambda t)\right]- \lambda \cos(\lambda t)\right) \vert}$,
 which shows the value $\frac{\tilde{R}(t) N a_B \sqrt{\vert \Omega(t) \vert} }{a_{\perp}}$ is  periodically varying with 
time but within the validity regime. So the validity criteria is satisfied for all values of time. Fig.~\ref{1d_app}(f) shows that the validity criteria is valid for all values of time for 
$ 
\tilde{R}(t)=1+\tanh\left(\frac{\Omega_0}{2} t\right)$ and $\vert \Omega(t)\vert =\sqrt{\frac{\Omega_0^2}{2}\left[1-\tanh\left(\frac{\Omega_0}{2} t\right)\right]}$. For Figs.~(\ref{1d_app}e) and ~(\ref{1d_app}f), see also Table 2.

\section{Solutions for the condition equations~(\ref{eq:cond}) for effective 1D GP equation mapped onto the NLS equation}\label{app:solve}
Substituting the relation $\Lambda(z,t)=r(t) \exp[i\theta(z,t)]$ from Eq.~(\ref{Lambda}) into the condition Eqs.~(\ref{eq:cond}) and separate them into real and imaginary parts, we get the following 
\begin{align}
r_t+\frac{1}{2} \Theta_{zz} r=0, \label{app_con:a}\\
\Theta_t+\frac{1}{2} \Theta_z^2+\frac{\Omega^2(t)}{2} z^2=0, \label{app_con:b}\\
F_t+\Theta_z F_z=0,  \label{app_con:c}\\
F_{zz}=0 \label{app_con:d}\\
G_t=\frac{F_z^2}{2}=r^2 \tilde{R} \label{app_con:e}
\end{align}
From Eqs.~(\ref{app_con:a}) and ~(\ref{app_con:d}), we obtain 
\begin{align}
\Theta(z,t)=& \frac{-r_t}{r} z^2+ \alpha_1(t) z+\alpha_z(t), \label{sol_con:a} \\
F(z,t)= & F_0(t) z+F_1(t), \label{sol_con:d} 
\end{align}
where $\alpha_1(t)$, $\alpha_2(t)$ and $F_1(t)$ are the functions of $t$ to be determined. 
Substituting the above form of the $\Theta(z,t)$ and $F(z,t)$ into Eqs.~(\ref{app_con:b}) and~(\ref{app_con:c}) and collecting the coefficient of the power $z$ and then solving them, we get the following 
\begin{align}
\alpha_1(t)= & b r^2 \label{sol_con:b1}\\
\alpha_2(t)= &-\frac{1}{2 }\int \alpha_1^2(t) dt \label{sol_con:b2}\\
F_0(t)= &\frac{1}{r_0} r^2 \label{sol_con:c1}\\
F_1(t)= &-\int \alpha_1(t) F_0(t) dt \label{sol_con:c2}\\
\frac{d}{dt}\left[ 2 \frac{r_t}{r}\right]- & \left[\frac{2 r_t}{r}\right]^2-\Omega^2(t)=0 \label{sol_con:b3} 
\end{align}
where $b$ and $r_0$ are the integrating constants.
From Eq.~(\ref{app_con:e})~ gives
\begin{align}
G(t)= & 2 \int r^2 \tilde{R} dt \label{app_con:e1}\\
F_0(t)= & r\sqrt{2 \tilde{R}} \label{sol_con:e2}
\end{align} 
Now comparing the Eqs.~(\ref{sol_con:c1}) and (\ref{sol_con:e2}), we get
\begin{align}
r= r_0 \sqrt{2 \tilde{R}}
\end{align} 

Substituting the above form of $r$ in to $\Theta$, $\alpha_1$, $\alpha_2$, $F_0$, $F_1$ and $G$ ,  we deduce the Eqs.~(\ref{solv}). Then Eq.~(\ref{con:eqn}) is obtained readily from the Eq.~(\ref{sol_con:b3}.) 

\section{General solutions to Ricatti equation}\label{app:ricatti}
Here we describe some of the interesting solutions of the restrictive  condition given by the Ricatti equation~(\ref{con:eqn}).
Substituting
\[ \tilde{R}=\exp\left[\int S(t)dt\right]
\]
in Eq.~(\ref{con:eqn}), we get a simplified form of the Ricatti equation as
\begin{align}\label{ricatti}
\frac{d S}{dt} -S^2-\Omega^2(t)=0.
\end{align}
  The general solution of the above Ricatti equation (\ref{ricatti}) can be readily given as ~\cite{Polyanin1995}
\begin{eqnarray}
S=S_0\frac{\exp[2\int S_0 dt]}{\left[\tilde{c}-\int\exp[2\int S_0 dt] dt \right]},
\end{eqnarray}
where $S_0$ is a particular solution of equation (\ref{ricatti}) and $\tilde{c}$ is an arbitrary constant.  From the general solution of (\ref{ricatti}), one can write down the general solution of  equation (\ref{con:eqn}) using
the relation $\tilde{R}=exp[\int S(t)dt]$.
There are numerous forms of $\Omega^2(t)$ for which explicit solutions for $\tilde{R}(t)$ can be obtained [see for example Ref.~\cite{Polyanin1995}]. Most important of them are given in the form of Table~\ref{table1}
\begin{table}[!ht]
\caption{Explicit solutions of the Riccati equation (\ref{con:eqn}) for different forms of
$\Omega^2(t)$.}
\label{table1} \centering
\begin{tabular}{|l|l|l|l|}
\hline  \multicolumn{2}{|c|}{Form of $\Omega^2(t)$}
	& \multicolumn{1}{c|}{Physically interesting}
	& \multicolumn{1}{c|}{General solution of $\tilde R$} \\
\multicolumn{2}{|c|}{}
	& \multicolumn{1}{c|}{particular solution of $\tilde R$}
	& \multicolumn{1}{c|}{($\tilde{c}$: arbitrary constant)} \\
\hline
1. & $\Omega_0^2$ = constant
	& $\sec(\Omega_0 t)$
	&  $\frac{1}{ \cos(\Omega_0 t)-  \tilde{c} \sin(\Omega_0 t)}$ \\ &&&\\
\hline
2. &$-\Omega_0^2$ = constant
	& $\sech(\Omega_0 t)$
	&  \\
   & &  &$\frac{1}{ \cosh(\Omega_0 t)- \tilde{c} \sinh(\Omega_0 t)}$ \\
   & & $\exp(\pm\Omega_0 t)$ & \\ &&&\\
\hline
3. & $-\frac{\Omega_0^2}{2}\left[1-\tanh\left(\frac{\Omega_0}{2} t\right)\right]$
	&  $1+\tanh\left(\frac{\Omega_0}{2} t\right)$
	&$\frac{2 \mbox{e}^{\Omega_0 t}}{-4 \tilde{c} +(1+\mbox{e}^{\Omega_0 t})-4 \tilde{c} (1+\mbox{e}^{\Omega_0 t}) \log(1+\mbox{e}^{\Omega_0 t})}$\\&& &\\
\hline
4.& $-\Omega_0\left(\frac{\Omega_0}{2}\left[1-\cos(2\lambda t)\right]+ \lambda \cos(\lambda t)\right)$ & $\exp\left[\frac{-\Omega_0}{\lambda}\cos(\lambda t)\right]$ &$\frac{\exp\left[\frac{-\Omega_0}{\lambda}\cos(\lambda t)\right]}{ 1- \tilde{c}\int \exp\left[\frac{-2\Omega_0}{\lambda}\cos(\lambda t)\right] dt}$\\&& &\\
\hline
5.& $\Omega_0 \lambda \exp(\lambda t)-\Omega_0^2 \exp(2 \lambda t)$
	& $\exp \left(\frac{\Omega_0}{\lambda} \exp(\lambda t)\right)$
	& $\frac{\exp \left(\frac{\Omega_0}{\lambda} \exp(\lambda t)\right)}{1-\tilde{c} \ExpIntegralEi \left(\frac{2\Omega_0}{\lambda} \exp(\lambda t)\right)}$\\&& &\\
\hline
6. &$3\Omega_0-\Omega_0^2 t^2$
	& $\frac{1}{t} \exp\left(\frac{\Omega_0t^2}{2}\right) $
	& $\frac{\exp\left(\frac{\Omega_0t^2}{2}\right)}{\tilde{c}\exp(\Omega_0t^2)+t-\sqrt{\Omega_0 \pi}\tilde{c} t \Erfi(\sqrt{\Omega_0} t)}$  \\& &&\\
\hline
7. &$\Omega_0-\Omega_0^2 t^2$ & $ \exp\left(\frac{\Omega_0t^2}{2}\right) $&
 $\frac{\exp\left(\frac{\Omega_0t^2}{2}\right)}{1+\sqrt{\pi}\tilde{c}
   \Erfi(\sqrt{\Omega_0} t)}$\\&& &\\
\hline
8.&$\Omega_0 n t^{n-1}-\Omega_0^2 t^{2n}$ & $ \exp\left(\frac{\Omega_0t^{n+1}}{n+1}\right) $ & Complicated integral \\
& & & \\
\hline
9.& $\frac{-2}{t^2}$&$\frac{3 t}{t^3+3 }$&$\frac{3 t}{\tilde{c}+3+t^3}$\\& &&\\
\hline
10.&$\frac{\tilde{b}}{t^2}, \tilde{b}< 0$ & $\frac{(2 \lambda+1)  t^{\lambda}}{(2 \lambda+1)  +t^{2 \lambda+1}}$, & Complicated integral \\
          & & where $\lambda=\frac{-1\pm \sqrt{1-4  \tilde{b}}}{2}$ &\\
\hline
11.& $\Omega_0 t \lambda -\Omega_0 (\Omega_0+ \lambda) \coth^2(\lambda t)$ & $ \left[\sinh(\lambda t)\right]^{\Omega_0 / \lambda}$& Complicated integral \\&& &\\
\hline
12.& $\Omega_0 t^2 (1-2 \coth^2(\Omega_0 t))$ & $\sinh(\Omega_0 t)$&$\frac{\sinh(\Omega_0 t)}{1-2 a \tilde{c} t+\tilde{c} \sinh(2 \Omega_0 t)}$\\
& && \\
\hline
13.& $ 3 \Omega_0 t \lambda -\lambda^2 -\Omega_0 (\Omega_0+ \lambda) \tanh^2(\lambda t)$ & $ \frac{\left[\cosh(\lambda t)\right]^{\Omega_0 / \lambda}}{\sinh(\lambda t)}$ & Complicated integral \\
& && \\
\hline
\end{tabular}
\end{table}
\addtocounter{table}{-1}
\begin{table}[!ht]
\caption{Continued.}
\label{table1a} \centering
\begin{tabular}{|l|l|l|l|}
\hline  \multicolumn{2}{|c|}{Form of $\Omega^2(t)$}  & Physically interesting $\tilde R$ & General solution of $\tilde R$ \\
\multicolumn{2}{|l|}{}  & Particular solution $\tilde R$ & \\
\hline 14.& $2 \Omega_0^2 \sech^2(\Omega_0 t)$ & $ \coth(\Omega_0 t)$&$\frac{\cosh(\Omega_0 t)}{(1-\Omega_0 \tilde{c} t)\sinh(\Omega_0 t)+\tilde{c} \cosh(\Omega_0 t)}$ \\&& &\\
\hline 15.& $ 3 \Omega_0 \lambda -\lambda^2 -\Omega_0 (\Omega_0+ \lambda) \coth^2(\lambda t)$ & $ \frac{\left[\sinh(\lambda t)\right]^{\Omega_0 / \lambda}}{\cosh(\lambda t)}$& Complicated integral \\&& &\\
\hline 16.& $-2 \Omega_0^2 \csech^2(\Omega_0 t)$ & $ \tanh(\Omega_0t)$& $\frac{\sinh(\Omega_0 t)}{(1-\Omega_0 \tilde{c} t)\cosh(\Omega_0 t)+\tilde{c}\sinh(\Omega_0 t)}$ \\ && &\\
\hline 17.& $\Omega_0 \lambda -\Omega_0 (a+ \lambda) \tanh^2(\lambda t)$ & $ \left[\cosh(\lambda t)\right]^{\Omega_0 / \lambda}$& Complicated integral \\&& &\\
\hline 18.& $\Omega_0^2 (1-2 \tanh^2(\Omega_0 t))$ & $ \cosh(\Omega_0 t)$&$\frac{\cosh(\Omega_0 t)}{1+2 \Omega_0 \tilde{c}t+\tilde{c} \sinh(2 \Omega_0 t)}$\\&& &\\
\hline 19.& $ -\Omega_0^2 +\Omega_0 \lambda \sinh(\lambda t)-\Omega_0^2 \sinh^2(\lambda t)$ & $ \exp\left(\frac{\Omega_0} {\lambda} \sinh(\lambda t)\right)$&$\frac{\exp\left(\frac{a }{\lambda}\sinh(\lambda t)\right)}{1-\tilde{c} \int \exp\left(\frac{2a }{\lambda}\sinh(\lambda t) dt\right)}$\\& &&\\
\hline 20.& $-2 \Omega_0^2 \left[\tanh^2(\Omega_0 t)+\coth^2(\Omega_0 t)\right]$ & $ \cosh(\Omega_0 t)\sinh(\Omega_0 t)$&$\frac{\cosh(\Omega_0t)\sinh(\Omega_0 t)}{1-4\Omega_0 \tilde{c} t+\tilde{c} \sinh(4 \Omega_0 t)}$\\&& &\\
\hline 21.& $ -\Omega_0^2 +\Omega_0 \lambda \cos(\lambda t)+ \Omega_0^2 \cos^2(\lambda t)$ & $ \exp\left(\frac{-\Omega_0 }{\lambda}\cos(\lambda t)\right)$&$\frac{\exp\left(\frac{-\Omega_0 }{\lambda}\cos(\lambda t)\right)}{1-\tilde{c} \int \exp\left(\frac{-2\Omega_0 }{\lambda}\cos(\lambda t) dt\right)}$\\&& &\\
\hline 22.&$ -\Omega_0^2 +\Omega_0 \lambda \sin(\lambda t)+ \Omega_0^2 \sin^2(\lambda t)$ & $
 \exp\left(\frac{-\Omega_0 }{\lambda}\sin(\lambda t)\right)$&
 $\frac{\exp\left(\frac{-\Omega_0 }{\lambda}\sin(\lambda t)\right)}
 {1-\tilde{c} \int \exp\left(\frac{-2\Omega_0 }{\lambda}\sin(\lambda t) dt\right)}$\\&& &\\
\hline 23.&$ \Omega_0 \lambda +\Omega_0 (\lambda-\Omega_0)\tan^2(\lambda t)$ & $ \sec(\Omega_0 t)^{\Omega_0/\lambda}$& Complicated integral \\& &&\\
\hline 24.& $ \lambda^2 +3 \Omega_0 \lambda+ \Omega_0(\lambda-\Omega_0)\tan^2(\lambda t)$ & $ \frac{\sec(\Omega_0 t)^{\Omega_0/\lambda}}{sin(\lambda t)}$& Complicated integral  \\&& &\\
\hline 25.&$ \lambda^2 +3 \Omega_0 \lambda+ \Omega_0(\lambda-\Omega_0)\cot^2(\lambda t)$ & $ \frac{\sec(\Omega_0 t)}{sin(\lambda t)^{\Omega_0/\lambda}}$& Complicated integral \\&& &\\
\hline 26.&$ \Omega_0 \lambda+ \Omega_0(\lambda-\Omega_0)\cot^2(\lambda t)$ & $\frac{1 }{sin(\lambda t)^{\Omega_0/\lambda}}$& Complicated integral \\&& &\\
\hline 27.&$-2\Omega_0^2 \left[\tan^2(\Omega_0 t)+\cot^2(\Omega_0 t)\right]$ & $ \cos(\Omega_0 t)\sin(\Omega_0 t)$&$\frac{\sin(2 \Omega_0 t)}{1- 4 \Omega_0 \tilde{c} t+ \tilde{c}\sin(4 \Omega_0 t)}$\\&& &\\
\hline
\end{tabular}
\end{table}


\section{Physical meaning of symbols used in the text}\label{app:symbol}

In order to make the various notations used in the text clear, we tabulate a list of symbols which we have used in the text with their physical meaning in Table \ref{tables} for ready reference.
\begin{table}[!ht]
\caption{Overview table of symbols used in the text and their physical meanings}
\label{tables} \centering
\begin{tabular}{|c|l|}
\hline  \multicolumn{1}{|c|}{symbol}
	& \multicolumn{1}{c|}{Physically meaning}  \\
\hline
 $\omega_{\perp}$
	& transverse trap frequency 	\\
\hline
$\omega_{z}$
	& axial trap frequency 	\\
\hline
$\Omega $
	& ratio of axial and transversive frequencies	$(\frac{\omega_z}{\omega_{\perp}})$\\
\hline
$\Gamma $
	& gain/loss of atoms	\\
\hline
$\gamma $
	& rescaled gain/loss of atoms  ($\frac{\Gamma}{\omega_{\perp}}$)\\
\hline
 $a_{\perp}$
	& transverse harmonic length 	\\
\hline
 $a_{B}$
	& Bohr radius 	\\
\hline
 $a_s$
	& s- wave scattering length 	\\
\hline
 $a_z$
	& longitudinal harmonic length \\
\hline
 $\zeta$
	& healing length \\
\hline
 $R$
	& rescaled s-wave scattering length   ($\frac{2 a_s}{a_B}$) \\
\hline
$\tilde{R}$
	& depends on s-wave scattering length and gain/loss term ($=\frac{2 a_s}{a_B} \exp \left[\int \gamma dt\right]$) \\
\hline
 $N$
	&number of the condensate atoms \\
\hline
 $n$
	&mean density of the condensate with cigar-shaped potential \\

\hline
 $m$
	&atomic mass of condensed atom \\
\hline
 $a$
	& amplitude of bright soliton of focusing NLS eqution \\
\hline
$\beta$
&depth of dark soliton of defocusing NLS eqution \\
\hline
$c$
	&velocity of soliton of NLS equation \\
\hline$b$,$r_0$ 
	&arbitrary integration constants \\
\hline
 $\sigma$
	&sign of nonlinearitiy ($\sigma=+1$ focusing, $\sigma=-1$ defocusing) \\
\hline
$ \Psi $
	& 3-D   GP equation wave function with gain/loss term	\\
\hline
$ \Phi $
	& effective 1-D   GP equation wave function with gain/loss term	\\
\hline
$ \Phi^+ $
	& generalized  bright soliton solution of the effective 1-D GP equation with gain/loss term	\\
\hline
$ \Phi^- $
	& generalized  dark soliton solution of the effective 1-D GP equation with gain/loss term\\

\hline
$ Q $
	& transformed effective 1-D  GP equation wave function 	\\
\hline

$ Q^+ $
	& generalized  bright soliton solution of the transformed effective 1-D GP equation	\\
\hline
$ Q^- $
	& generalized  dark soliton solution of the transformed effective 1-D GP equation\\

\hline
$ q $
	& 1-D NLS equation wave function\\
\hline
$ q^+ $
	& bright soliton of NLS equation	\\
\hline
$ q^- $
	&  dark soliton	of NLS equation\\
\hline
$\xi^+$
	& wave variable which specifies the width and velocity of the bright soliton solution of effective 1-D GP equation \\
\hline
$ \xi^- $
	& wave variable which specifies the  width and velocity of the dark soliton solution of effective 1-D GP equation  \\
\hline
$ \eta^+ $
	& phase of bright soliton solution of effective 1-D GP equation	with gain term\\
\hline
$ \eta^- $
	&  phase of dark soliton solution of effective 1-D GP equation with gain term\\
\hline
$ A^+ $
	& amplitute of bright soliton solution of effective 1-D GP equation with gain term\\
\hline
$ A^- $
	& amplitute of dark soliton solution of effective 1-D GP equation with gain term\\
\hline
\end{tabular}
\end{table}


\end{document}